\documentclass[twocolumn]{aastex62}

\usepackage{dot2texi}
\usepackage{tikz}
\usetikzlibrary{bayesnet}
\tikzset{Bayes net/.code=\BayesTemp}

\usepackage{graphicx}
\usetikzlibrary{arrows}
\usepackage{amsmath,gensymb,xcolor}
\accepted{for publication in ApJ}

\shorttitle{REQUIEM-2D: Spatially Resolved Stellar Populations from \emph{HST} 2D Grism Spectroscopy}
\shortauthors{M. Akhshik, K. Whitaker, G. Brammer et. al.}

\begin{document}

\title{REQUIEM-2D Methodology: Spatially Resolved Stellar Populations of Massive Lensed Quiescent Galaxies from Hubble Space Telescope 2D Grism Spectroscopy}

\correspondingauthor{Mohammad Akhshik}
\email{mohammad.akhshik@uconn.edu}

\author[0000-0002-3240-7660]{Mohammad Akhshik}
\affil{Department of Physics, University of Connecticut,
Storrs, CT 06269, USA}

\author[0000-0001-7160-3632]{Katherine E. Whitaker}
\affil{Department of Astronomy, University of Massachusetts, Amherst, MA 01003, USA}
\affil{Cosmic Dawn Center (DAWN)}

\author[0000-0003-2680-005X]{Gabriel Brammer}
\affil{Cosmic Dawn Center (DAWN)}
\affil{Niels Bohr Institute, University of Copenhagen, Lyngbyvej 2, DK-2100 Copenhagen, Denmark}

\author[0000-0003-3266-2001]{Guillaume Mahler}
\affil{Department of Astronomy, University of Michigan, 1085 South University Ave, Ann Arbor, MI 48109, USA}

\author[0000-0002-7559-0864]{Keren Sharon}
\affil{Department of Astronomy, University of Michigan, 1085 South University Ave, Ann Arbor, MI 48109, USA}

\author[0000-0001-6755-1315]{Joel Leja}
\affil{NSF Astronomy and Astrophysics Postdoctoral Fellow}
\affil{Harvard-Smithsonian Center for Astrophysics, 60 Garden St. Cambridge, MA 02138, USA}

\author[0000-0003-1074-4807]{Matthew B. Bayliss}
\affil{Department of Physics, University of Cincinnati, Cincinnati, OH 45221, USA}

\author[0000-0001-5063-8254]{Rachel Bezanson}
\affil{Department of Physics and Astronomy, University of Pittsburgh, Pittsburgh, PA 15260, USA}

\author[0000-0003-1370-5010]{Michael D. Gladders}
\affil{Department of Astronomy \& Astrophysics, The University of Chicago, 5640 South Ellis Avenue, Chicago, IL 60637, USA}
\affil{Kavli Institute for Cosmological Physics at the University of Chicago, 5640 South Ellis Avenue, Chicago, IL 60637, USA}

\author[0000-0003-2475-124X]{Allison Man}
\affil{Dunlap Institute for Astronomy and Astrophysics, University of Toronto, 50 St George Street, Toronto ON, M5S 3H4, Canada}

\author[0000-0002-7524-374X]{Erica J. Nelson}
\affil{Hubble Fellow}
\affil{Harvard-Smithsonian Center for Astrophysics, 60 Garden St. Cambridge, MA 02138, USA}

\author[0000-0002-7627-6551]{Jane R. Rigby}
\affil{Observational Cosmology Lab, NASA Goddard Space Flight Center, 8800 Greenbelt Rd., Greenbelt, MD 20771, USA}

\author{Francesca Rizzo}
\affil{Max-Planck Institute for Astrophysics, Karl-Schwarzschild Str 1, D-85748 Garching, Germany.}

\author[0000-0003-3631-7176]{Sune Toft}
\affil{Cosmic Dawn Center (DAWN)}
\affil{Niels Bohr Institute, University of Copenhagen, Lyngbyvej 2, DK-2100 Copenhagen, Denmark}

\author[0000-0002-3977-2724]{Sarah Wellons}
\affil{Center for Interdisciplinary Exploration and Research in Astrophysics (CIERA) and Department of Physics and Astronomy, Northwestern University, Evanston, IL 60208, USA}

\author[0000-0003-2919-7495]{Christina C. Williams}
\affil{NSF Astronomy and Astrophysics Postdoctoral Fellow}
\affil{Steward Observatory, University of Arizona, 933 North Cherry Avenue, Tucson, AZ 85721, USA}

\begin{abstract}

We present a novel Bayesian methodology to jointly model photometry and deep \emph{Hubble Space Telescope} (\emph{HST}) 2d grism spectroscopy of high-redshift galaxies. Our \texttt{requiem2d} code measures both unresolved and resolved stellar populations, ages, and star-formation histories (SFHs) for the ongoing REQIUEM (REsolving QUIEscent Magnified) Galaxies Survey, which targets strong gravitationally lensed quiescent galaxies at z$\sim$2. We test the accuracy of \texttt{requiem2d} using a simulated sample of massive galaxies at z$\sim$2 from the Illustris cosmological simulation and find we recover the general trends in SFH and median stellar ages. We further present a pilot study for the REQUIEM Galaxies Survey: MRG-S0851, a quintuply-imaged, massive ($\log M_* / M_\odot = 11.02\pm 0.04$)  red galaxy at $z=1.883\pm 0.001$. With an estimated gravitational magnification of $\mu = 5.7^{+0.4}_{-0.2}$, we sample the stellar populations on 0.6 kpc physical size bins. The global mass-weighted median age is constrained to be $1.8_{-0.2}^{+0.3}$ Gyr, and our spatially resolved analysis reveals that MRG-S0851 has a flat age gradient in the inner 3 kpc core after taking into account the subtle effects of dust and metallicity on age measurements, favoring an early formation scenario.  The analysis for the full REQUIEM-2D sample will be presented in a forthcoming paper with a beta-release of the \texttt{requiem2d} code.

\end{abstract}


\section{Introduction} \label{sec:intro}

Our understanding of galaxies a few billion years after the Big Bang has dramatically improved over the last few decades. It is now well established that galaxies follow a bi-modal color distribution in both the low and high redshift universe \cite[e.g.,][]{strateva2001,whitaker2011}, including a population of old, red, more massive quiescent galaxies and a population of young, blue, less massive star-forming galaxies.  Star-forming and quiescent galaxies can be identified by their location in the star formation rate (SFR) versus stellar-mass plane, where star-forming populations form a sequence with a relatively low scatter \citep[e.g.,][]{whitaker2014,speagle2014}; and quiescent populations lie well below the average relation. 
The number density of massive quiescent galaxies rapidly increased at early times, comprising
up to half of the total massive galaxy population by $z\sim 2$ \citep{kriek2006,brammer2011,muzzin2013}.  Moreover, observations show these quiescent galaxies to be remarkably compact relative to star-forming galaxies with similar stellar masses at a given redshift \citep[e.g.,][]{vandokkum2008,vanderwel2014}, with only the most massive galaxies ($\log M_* / M_\odot > 11.3$) having similar size distributions as the star-forming population \citep{mowla2018}. 

Despite the tremendous progress in understanding the population of z$\sim$2 massive galaxies, usually presented in empirical correlations like the SFR-stellar mass correlation of star-forming galaxies described above, the physical mechanism(s) responsible for quenching star-forming galaxies remain unknown. Spatially resolved spectroscopy and imaging hold the power to address these fundamental questions. Simulations suggest that stellar age and specific star-formation rate gradients can constrain the theoretical formation scenarios for high-redshift massive quiescent galaxies \citep[e.g.,][]{wellons2015,Tacchella2015,tacchella2016}. However, the low spatial resolution of near and mid-infrared imaging and the high stellar-density of quiescent galaxies mostly limit the studies to the spatially-unresolved data with relatively less constraining power to distinguish between theoretical models \citep[e.g.,][]{williams2017,abramson2018,belli2019,estrada-carpenter2020}. Strong gravitational lensing offers a solution for this challenge as it magnifies distant galaxies and boosts their signal-to-noise ratio (SNR).  Furthermore, the actual un-lensed morphology can be reconstructed accurately with a detailed lensing model \citep[e.g.,][]{sharon2012,sharon2020}. 

Strong gravitationally-lensed galaxies are discovered and studied extensively in the literature \citep[e.g.,][]{williams1996, yee1996, allam2007, smail2007, siana2008, belokurov2009, lin2009, koester2010, sharon2012, gladders2013}, with many cases of spatially resolved stellar population analyses in star-forming galaxies \cite[e.g.,][]{stark2008, swinback2009, jones2010, leethochawalit2016}. Despite their rarity, a number of ground-based spectroscopic studies of massive quiescent galaxies have steadily accumulated within the literature \citep[e.g., Keck/MOSFIRE, Magellan/FIRE, and VLT/X-Shooter;][]{muzzin2012,geier2013,newman2015,hill2016,toft2017,newman2018,newman2018b,ebeling2018}.  However, ground-based spatial resolution is insufficient to resolve spectroscopic signatures of the stellar populations of all but perhaps the most strongly lensed objects \citep{newman2015}. 

The high spatial and low spectral resolution of grism spectroscopy with the \emph{HST}/Wide Field Camera 3 (WFC3) enables measuring both the unresolved and/or resolved stellar populations \citep[e.g.,][]{vandokkum2010a,brammer2012a,whitaker2013,whitaker2014a,estrada2019,estrada-carpenter2020,deugenio2020}. In particular, \citet{abramson2018} use WFC3/G141 grism spectroscopy and multi-wavelength \emph{HST} imaging to study the spatially resolved stellar populations of four massive galaxies at z$\sim$1.3, finding a link between bulge mass function and the shape of the star-formation history.  Similar comprehensive studies of massive quiescent galaxies at higher redshifts demonstrate that it is feasible to reconstruct SFHs based on a joint spectro-photometric \emph{HST} analyses \citep{morishita2018,morishita2019}.  While the measured metallicities of quiescent galaxies at $z\sim2$ are generally found to be similar to local early-type galaxies \citep{morishita2019}, there exist a few lensed quiescent galaxies with lower metallicities that suggest a mechanism other than dry minor-mergers would be necessary to explain their chemical enrichment \citep{morishita2018}.

In this paper, we present our methodology developed to jointly fit \emph{HST} and \emph{Spitzer}-IRAC spectro-photometric data in preparation for the analysis of the full REQUIEM galaxy survey (HST-GO-15633). While it is possible to constrain stellar population properties by analyzing spatially-resolved\footnote{We caution that the term spatially ``resolved'' in nearby galaxies is  reserved for an observational study that can resolve stars down to at least $\lessapprox \mathcal{O}({10^6})$ stars per pixel \citep[e.g.,][]{cook2019}. This limit corresponds to distances of $<$1 Mpc with \emph{HST} detectors, and nominally, even individual stellar clusters can be identified in ``spatially resolved'' studies for nearby targets and surveys \citep[e.g.,][]{johnson2012,johnson2015}. Our targets are well beyond this limit, but we can still resolve stellar populations down to a fraction of kpc scale, and we therefore use the term spatially ``resolved'' to refer to our study, noting the conceptual difference in the terminology used for nearby and z$\sim$2 galaxies.} spectroscopic and photometric data separately, we perform a joint spectro-photometric fit, since using a joint fit, we can optimally use all spectro-photometric data \citep[e.g.,][]{newman2014} and infer all parameters within a single framework. As a single set of assumptions is applied in this joint-fitting, it is also easier to understand and address potential biases and systematics.

We briefly introduce the REQUIEM galaxy survey in Section \ref{sec:req-survey}, and illustrate our method using a pilot target, MRG-S0851 \citep{sharon2020}. The \emph{HST} and \emph{Spitzer} data reductions are presented in Section \ref{sec:data}. The methodology to jointly fit photometry and spectroscopy is presented in Section \ref{subsec:method}.  We discuss inferring ages and star-formation histories in Section \ref{sec:age-appendix}, testing the inferred parameters using a sample of massive quiescent and star-forming galaxies selected from the Illustris simulation. In Section \ref{subsec:real-test}, we present first results from REQUIEM-2D grism spectroscopy for our pilot target, MRG-S0851. In Appendix \ref{sec:cluster-lens} and \ref{sec:whole arc}, we discuss the details of the lensing model and the morphological measurements of MRG-S0851.

In this paper we adopt a standard simplified $\mathrm{\Lambda CDM}$ cosmology with $\Omega _{M}=0.3$, $\Omega _{\Lambda}=0.7$ and $H_0 = 70 \, \mathrm{km}/\mathrm{s}/\mathrm{Mpc}$. We assume the \citet{chabrier2003} initial mass function (IMF). All magnitudes are reported in the AB system.

\begin{figure*}[!t]
\centering
\includegraphics[width=0.9\textwidth]{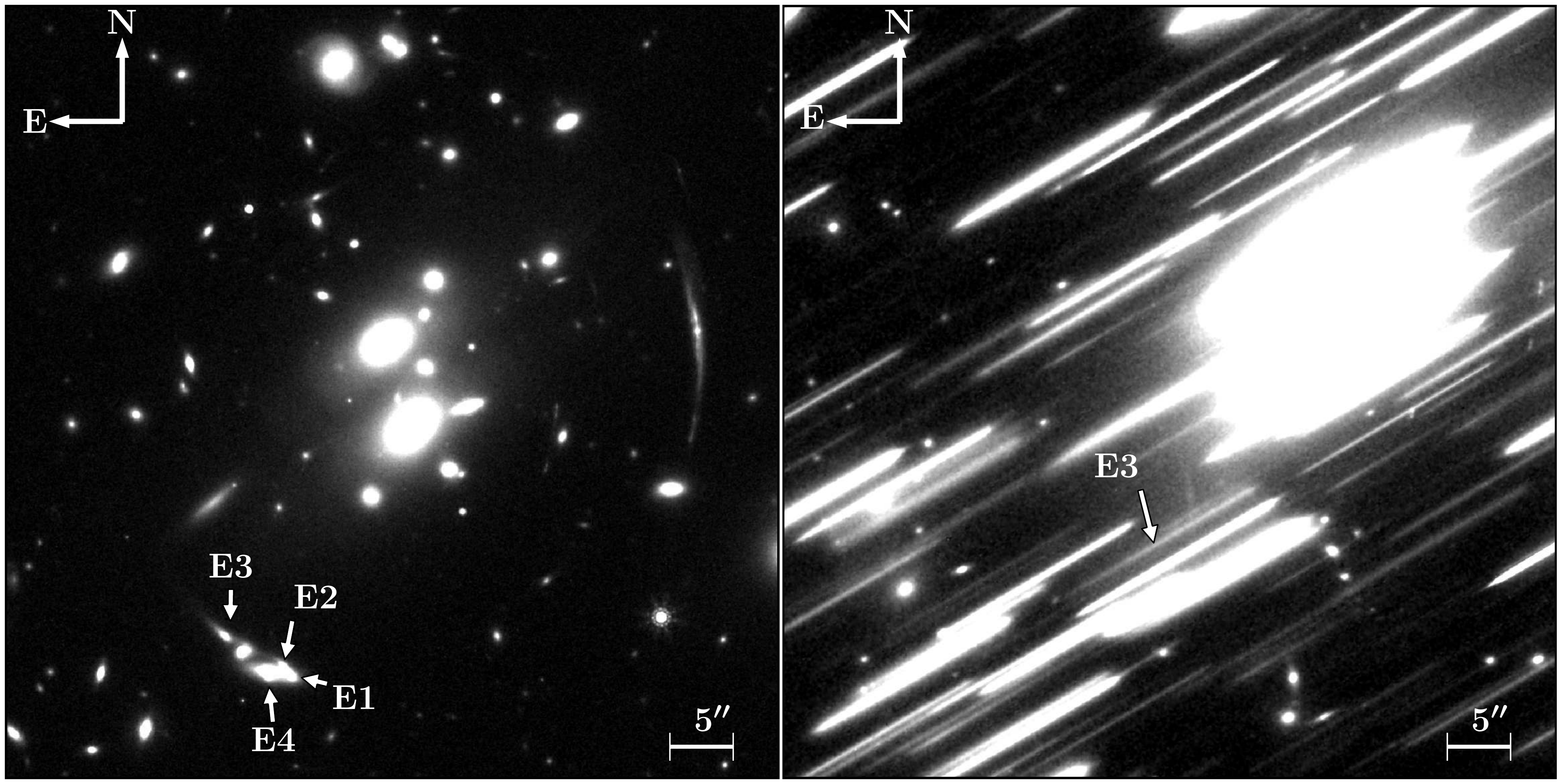}
\caption{The \emph{HST} imaging and spectroscopy of MRG-S0851. On the left, the drizzled mosaics of WFC3 $H_{\mathrm{F160W}}$ filter is shown. The four main images of MRG-S0851, a pilot target from REQUIEM survey, are indicated by white arrows. The fifth image, E5, is right next to the subcluster lens and it is not shown here. We indicate E5 in Figure~\ref{fig:0851-E-galfit}, where the light profile of subcluster lens is modeled and subtracted from the image.  On the right, the drizzled mosaic for $\sim 6$ orbits of the WFC3/G141 data is shown. White arrow indicates the grism spectrum of E3. E3 is the cleanest and the brightest image of the system, MRG-S0851 in the H-band, with our \emph{HST} grism observations optimized to reduce contamination for E3 at the expense of losing E1, E2 and E4. \label{fig:0851-E-mosaics}}
\end{figure*}

\section{REQUIEM-2D Galaxy Survey \label{sec:req-survey}}

Capitalizing on the decade-long hunt for strong lensed quiescent galaxies at $z>1.5$ and the slitless spectroscopic capabilities of \emph{HST}, the REQUIEM-2D galaxy survey targets 8 strongly lensed quiescent galaxies spanning redshifts of $1.6<z<2.9$, stellar masses of $10.4<\log M_*/M_\odot<11.7$, and specific star formation rates of $\log \mathrm{sSFR}_{\mathrm{100 Myr}}/[\mathrm{yr^{-1}}]<-10.3$ (HST-GO-15633)

 Next we briefly introduce the targets comprising the REQUIEM-2D survey, with the pilot target MRG-S0851 described in further detail in Section \ref{subsec:real-test}.  Our sample includes: 
 
 \begin{itemize}
     \item MRG-M1341: a highly magnified $\mu\sim 30$ galaxy at $z=1.6$ \citep{ebeling2018} (15 orbits of WFC3/G141),
     \item MRG-S0851: A massive lensed red galaxy at $z$=1.88, with centrally-concentrated rest-frame UV flux (12 orbits of WFC3/G141; presented in this paper)
    \item MRG-M0138: a massive and bright target at $z=1.95$ with $\log M/M_\odot$=11.7 and $\mathrm{H_{F160W}}=17.3$ \citep{newman2018} (6 orbits of WFC3/G141),
    \item MRG-P0918 and MRG-S1522, relatively young quiescent galaxies at $z=2.36$ and $z=2.45$, respectively, with ages of 0.5-0.6 Gyr \citep{newman2018} (7 orbits of WFC3/G141 each),
    \item MRG-M2129, a rotationally-supported quenched galaxy at $z=2.1$ \citep{toft2017} (5 orbits of WFC3/G141),
    \item MRG-M0150, a dispersion-dominated ($V/\sigma=0.7\pm0.2$) massive quiescent galaxy at $z=2.6$ \citep{newman2015} (5 orbits of WFC3/G141), and 
    \item MRG-S0454, the most compact $r_{\mathrm{eff}}\sim 0.3$kpc target of the REQUIEM-2D survey with the highest redshift of $z=2.9$ (Man et. al. in prep) (12 orbits of WFC3/G141).
  \end{itemize}

The number of \emph{HST} bands available for the REQUIEM targets ranges from a minimum of 5 filters to a maximum of 16.  All targets have photometric coverage from $\sim 1000\mathrm{\AA}$ to $\sim 15000 \mathrm{\AA}$ in rest-frame wavelength, and grism G141 coverage varies from rest-frame wavelengths of $\sim 2900-4200 \mathrm{\AA}$ for the target with the highest redshift to $\sim 4400-6300 \mathrm{\AA}$ for the target with the lowest redshift. The imaging data used herein for the test target, MRG-S0851, consists of 5 \emph{HST} bands and 2 \emph{Spitzer} bands (see Section~\ref{subsec:real-test}).

\section{Data Reduction and Analysis} \label{sec:data}

\subsection{Hubble Space Telescope Grism Spectroscopy}

The REQUIEM-2D \emph{HST} observations are designed following the 3D-HST standard \citep{brammer2012}, including a shorter exposure with a WFC3/IR imaging filter immediately before/after two longer WFC3/IR G141 exposures. The ``Grism redshift \& line analysis software for space-based slitless spectroscopy'', or \texttt{Grizli}, is used for the data reduction analysis \citep{brammer2016}. \texttt{Grizli} is specifically designed for manipulating \emph{HST} slitless spectroscopic observations and serves for the data reduction herein.

Astrometric calibrations of the WFC3-IR and WFC3-UVIS images are performed in two steps within \texttt{Grizli}. In the first step, the relative astrometry is set by aligning all available exposures in each filter together.  The Pan-STARSS catalog \citep{flewelling2016} is then used to density match the detected objects. The absolute astrometric registration is finally improved by adopting the Gaia-DR2 catalog \citep{gaia2018a,gaia2018b}.

\texttt{Grizli} matches the world coordinate system (WCS) of the grism exposures with already-registered WFC3/IR exposures and subtracts the sky background after reducing and calibrating the grism exposures. All of the exposures are drizzled together using the \texttt{AstroDrizzle} package \citep{avila2012}. Figure~\ref{fig:0851-E-mosaics} shows the final product for MRG-S0851, a pilot target from the REQUIEM galaxy survey (see Section \ref{subsec:real-test}). 

\begin{figure}
\centering
\includegraphics[width=1.0\columnwidth]{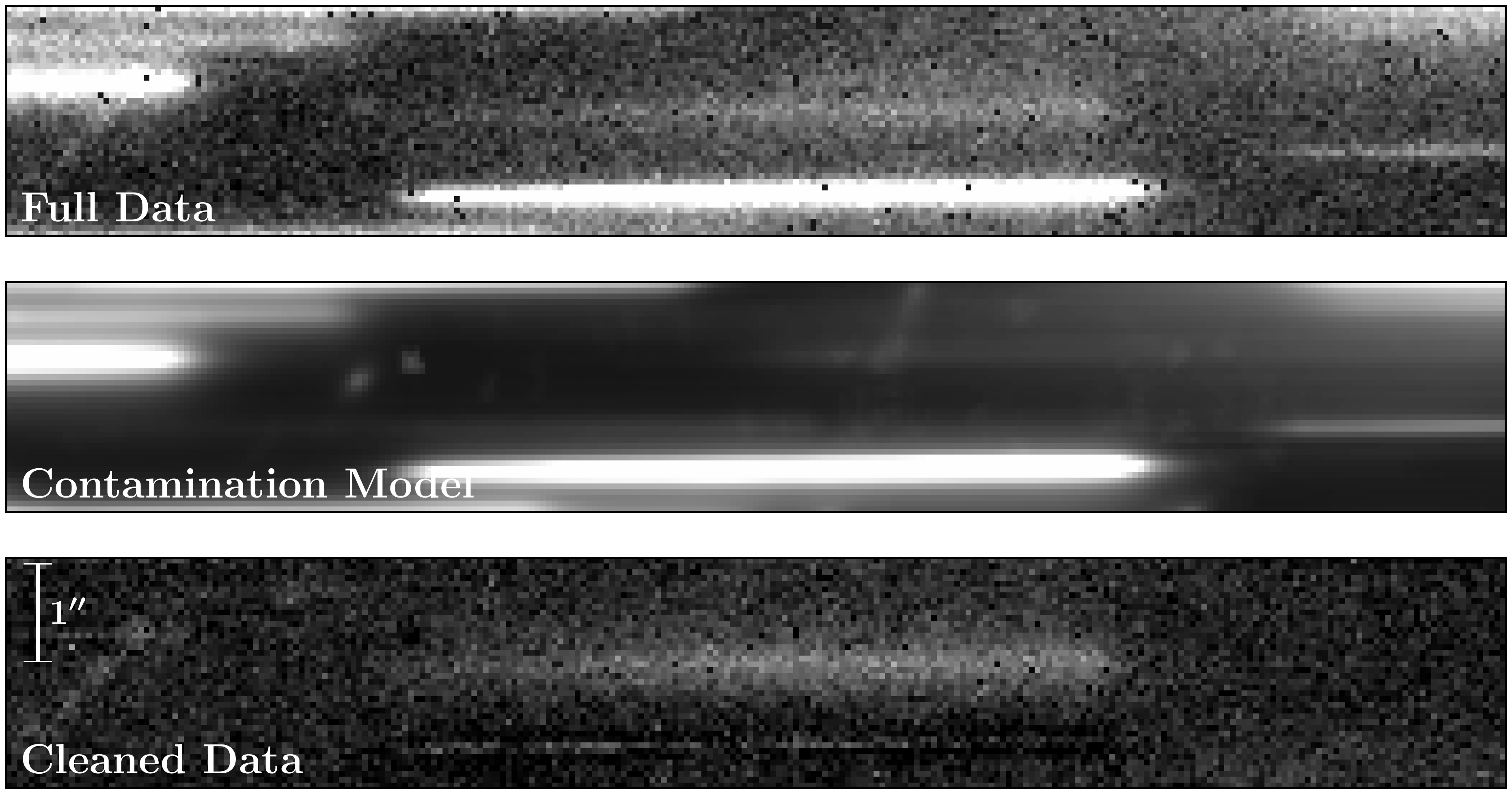}
\caption{A $15\farcs6 \times 2\farcs4$ cutout region of a single G141 exposure centered at E3, the brightest image of the pilot target MRG-S0851. In the top panel, we show the full data. The second panel shows the contamination model, which we obtain by iteratively fitting polynomial spectral templates to the grism spectra of all objects, and in the third panel, we show cleaned grism data of E3. \label{fig:2d-grism}}
\end{figure}

WFC3/G141 grism produces dispersed spectra of every object within the field-of-view of the instrument. Without slits, however, spectra of nearby objects overlap. To analyze the 2D grism spectrum of an object of interest, contamination by other objects must be removed. Here, we adopt an iterative algorithm within \texttt{Grizli} to remove contamination. First, 2D grism models are generated for all objects assuming a flat spectrum that we refine iteratively. In subsequent steps, we concentrate on the region surrounding the primary science target, which extends roughly a factor of 5 times beyond the largest spatial extent of the main science target. The grism model of all objects is refined in this surrounding region by using a second and fifth degree polynomials as spectral templates. A linear combination of a flexible set of spectral templates, that are built in \texttt{Grizli} to constrain redshifts \citep{brammer2008}, is finally fit to improve the quality of the model. Figure \ref{fig:2d-grism} demonstrates the result of this procedure for the pilot target, MRG-S0851.

\subsection{Photometric Measurements}

To perform the joint spectro-photometric analysis, we first construct a photometric catalog, largely following \citet{whitaker2011} and \citet{skelton2014}.  We refer the reader to these papers for a more in-depth discussion on the methodology adopted.

\subsubsection{Hubble Space Telescope Photometry}

To detect sources, we first construct a noise-equalized image by multiplying the $\mathrm{H}_{\mathrm{F160W}}$ mosaic with the square root of the corresponding weight map. We then run \texttt{Source Extractor} \citep{bertin1996} on this image. The detection threshold is set at $1.8\sigma$, the deblending threshold at 32, with a minimum contrast of 0.0001, and a minimum area of 14 pixels.

To create the point-spread functions (PSF), a stellar sequence is identified by considering the ratio of a small aperture ($0\farcs5$) flux to a large aperture ($2^{\prime\prime}$) flux for each band. Stars form a tight sequence close to unity, making them easily identifiable above a certain threshold in magnitude.  A $5^{\prime\prime}$ postage stamp cutout of each bright star is created. An average PSF is calculated after centering and normalizing the stamps. The PSF matching is performed using a kernel that convolves each PSF to match the $\mathrm{H}_{\mathrm{F160W}}$ PSF as a reference, since it has the largest full width at half maximum (FWHM) of $0\farcs18$. To obtain the kernel, we use custom codes that fit a set of Hermite polynomials weighted by Gaussian two dimensional profiles to the Fourier transform of the stacked stars. The PSF homogenization is accurate within a percent level.

Next, \texttt{Source Extractor} is run in the dual-image mode with the noise-equalized image of $\mathrm{H}_{\mathrm{F160W}}$ as a detection image and the PSF-matched mosaic of interest as a measurement image, including the weight maps of the PSF-matched mosaics as well. The photometry is calculated adopting an aperture of $1\farcs5$ diameter for all but the most extended strong lensed sources.  This is about a factor of two larger than the aperture size adopted in earlier works, but justified when analyzing strong gravitationally lensed sources with linear magnifications of $\mu\sim3-6$ (e.g., $0\farcs7\sim1\farcs5/\sqrt{\mu}$), since the larger aperture in the image plane of lensed target effectively covers the same physical region in the source plane as the smaller aperture would cover for unlensed targets.  

The curve of growth of the $\mathrm{H}_{\mathrm{F160W}}$ filter is used to correct the \texttt{AUTO} flux value reported by \texttt{Source Extractor} for the amount of light falling outside the Kron radius \citep{kron1980}. This correction factor is the ratio of the total flux of a point source in $\mathrm{H}_{\mathrm{F160W}}$ to the the flux enclosed in the Kron radius \citep[e.g.,][]{skelton2014}.

Realistic uncertainties are estimated by placing apertures in empty regions across the image and calculating the noise properties directly from the images in lieu of using the standard weight maps, noting that the drizzling process correlates the pixels, and as a result the uncertainty inferred from the weight maps is underestimated \citep[e.g.][]{casertano2000}. More details can be found in Section 3.5 of \citet{whitaker2011} and Section 3.4 of \citet{skelton2014}.

\subsubsection{Spitzer/IRAC Photometry}

To obtain photometric measurements from the low resolution \emph{Spitzer} observations, we use the Multi-resolution Object PHotometry ON Galaxy Observations code \citep[\texttt{MOPHONGO};][]{labbe2006,wuyts2007}. \texttt{MOPHONGO} makes two dimensional models for different objects in the field and uses them to deblend and measure fluxes, taking into account the difference in PSF between \emph{Spitzer} and \emph{HST} images.

Following \citet{whitaker2011}, \emph{Spitzer} photometric fluxes are measured using $3''$ diameter apertures size, applying photometric corrections using the $\mathrm{H}_{\mathrm{F160W}}$ curve of growth. While poor resolution, the photometric measurements of the two \emph{Spitzer} IRAC channels play a crucial role in the modeling of stellar populations owing to the extended wavelength coverage into the rest-frame near-infrared at $z\sim2$ that helps to constrain the dust. \citep[for example see,][]{muzzin2008}. 

\section{Methodology to Fit the Age and SFH of the Stellar Populations} \label{subsec:method}

 In this Section, we discuss the methodology used by the \texttt{requiem2d} software package to combine all spectroscopic and photometric data and constrain the age and SFHs of unresolved and resolved stellar populations. An overview of the main aspects of our methodology is presented in Section \ref{subsec:method-lite}.  We then outline our approach to model dust and metallicity in Section \ref{subsec:dust-metal-prior}, before formally introducing the elements of the full model in Section \ref{subsec:model-elements}. A discussion on priors and the computational Bayesian approach can be found in Section \ref{subsec:priors}.

\subsection{Overview of Methodology \label{subsec:method-lite}}

The \texttt{requiem2d} package adopts a non-parametric framework to model SFHs, avoiding any assumptions about their functional form (see Section \ref{subsec:priors} for a discussion of SFH priors). Joint spectro-photometric fitting is particularly important for a robust analysis of the stellar populations, with the longer wavelength baseline of the photometry helping to constrain dust and the higher spectral resolution grism spectroscopy providing more robust constraints on redshift, age, and metallicity by constraining spectral absorption lines.

We adopt a non-parametric approach to analyze SFHs, specifically modeling the composite stellar population (CSP) of the targets as a linear combination of simple stellar populations (SSPs) \citep[e.g.,][]{heavens2004,ocvirk2006,panter2007,tojeiro2007,kelson2014,leja2017,dressler2018,morishita2019}, which is used to constrain the ``weights'' of each SSP, denoted herein by $\mathbf{x}$. The secondary parameters such as age and star-formation rate (SFR) are then calculated using these weights. This methodology, in principle, is similar to the approach adopted in EAZY \citep{brammer2008}, where one fits a linear combination of templates to photometric data to constrain the redshift. Here, we fit a linear combination of the SSP templates with varying ages to the low-resolution spectroscopic and photometric data.

To generate SSPs, we use the dust and metallicity posteriors obtained by fitting the photometric data alone (Section~\ref{subsec:dust-metal-prior}). We then refit the full spectro-photometric data to infer ages and SFHs using these SSPs (Section~\ref{subsec:model-elements}). In the remainder of this Section, we discuss data preparation steps (Sections~\ref{subsec:def-bins} and \ref{subsec:data-prep}).

\subsubsection{Defining the Spatial Bins \label{subsec:def-bins}}

To study the spatially resolved stellar populations for lensed targets, we define spatial bins for each grism exposure separately, using the corresponding direct WFC3/IR image with the same pixel scale and orientation as the grism exposure. ``Rows of pixels" are defined parallel to the dispersion angle $P_\theta$. We identify the row which has the pixel with the highest flux in the image and add two adjacent pixel rows to define the central bin. 

On either side of the center bin, two bins are defined that are 3-4 pixel rows wide respectively. Depending on the magnification of the main science target, either new subsequent spatial bins are added, or the rest of the pixels on each side are grouped to define the final outer bins. These other bins include the pixels corresponding to the low SNR portion of the extended light profiles. 

With a pixel size of $0\farcs06$, the central bins range from $0\farcs18$ to $0\farcs24$ wide. Lens models are used to determine the source-plane position of the defined spatial bins. For our pilot study of MRG-S0815 (Section \ref{subsec:real-test}), we probe the age gradient in the inner radius of $\sim$1.8$\, \mathrm{kpc}$ at an average spatial resolution of $\sim 0.6 \, \mathrm{kpc}$.

\subsubsection{Preparing the Data \label{subsec:data-prep}}

Grism spectra are analyzed in the native 2D space, limiting to grism pixels with a minimum SNR of 3. We also only include the grism pixels with less than 10\% contamination by adjacent objects.

It is not trivial to spatially-resolve the \emph{Spitzer}/IRAC bands. In our final joint-fitting, we therefore adopt a conservative approach by requiring that the total IRAC fluxes of all spatial bins match the global measured IRAC flux of the object. We discuss other complications of not having resolved rest-frame near-infrared (IRAC) fluxes in Section \ref{subsec:joint-fit-prior}.

The grism spectra of the bins overlap in 2D space, making it impossible to extract the 2D grism spectrum for each bin individually unless we have the best model for the other bins. We therefore construct a model for each bin individually with \texttt{Grizli}, then add all of the bins' models together to get a model for the whole galaxy. 

We could in principle model all grism exposures individually and compare them with the observed grism exposure. However, to reduce the computational cost, we use drizzled grism images in our analyses, constructed by combining all grism exposures of each dispersion angle. 

\begin{table*}[!t]
    \centering
    \begin{tabular}{|c|c|c|c|c|c|c|}
         \hline Fit & Software & SSP Prior & Grism & Res \emph{HST} Phot & Unres \emph{HST} Phot & \emph{Spitzer} Phot  \\ \hline
         Global Phot & \texttt{Prospector-$\alpha$} & $\times$ & $\times$ & 
         $\times$ & $\checkmark$ & $\checkmark$ \\ \hline
         Resolved Phot & \texttt{Prospector-$\alpha$} & $\times$ & $\times$ & $\checkmark$ & $\times$ & $\times$  \\ \hline
         Global Spec+Phot & \texttt{requiem2d} & Global Phot & $\checkmark$ & $\times$ & $\checkmark$ & $\checkmark$ \\ \hline 
         Resolved Spec+Phot & \texttt{requiem2d} & Global Phot & $\checkmark$ & $\checkmark$ & $\times$ & $\checkmark$ \\ \hline 
        Resolved Spec+Phot & \texttt{requiem2d} & Resolved Phot & $\checkmark$ & $\checkmark$ & $\times$ & $\checkmark$ \\ \hline

    \end{tabular}
    \caption{Table of 5 different fits performed in our analyses, indicating a software used, included data and SSP prior if it is used. Phot is a shorthand for photometry, Spec stands for spectroscopy, and Res and Unres stand for Resolved and Unresolved respectively.}
    \label{tab:fits}
\end{table*}
\subsection{Measuring Dust and Metallicity \label{subsec:dust-metal-prior}}

The main goal when using \texttt{requiem2d} is to constrain the ages and star-formation histories of massive quiescent galaxies at $z\sim2$, treating dust and metallicity as nuisance parameters. While both of these parameters are degenerate with age, they are not well-constrained by the relatively short wavelength coverage and low spectral resolution of grism spectroscopy. Hence, our strategy is to analyze the problem in two steps: 

\begin{enumerate}
    \item Photometric data are fitted alone using \texttt{Prospector-$\alpha$} \citep{leja2017} to obtain the posterior of dust, metallicity and other relevant parameters of stellar populations such as the stellar-mass (Section \ref{subsec:prospector-phot-fit}), and the posteriors of dust and metallicity are subsequently used to generate SSPs (Section \ref{subsec:joint-fit-prior}).
    \item  Joint fit of photometric and spectroscopic data are preformed using \texttt{requiem2d} code to constrain the age and SFHs of the stellar populations, using the SSPs generated at the first step \textbf{(Section \ref{subsec:model-elements})}.
\end{enumerate}

\subsubsection{\texttt{Prospector-$\alpha$} Fit to Resolved and Global Photometric Data \label{subsec:prospector-phot-fit}}
Following the same steps and assumptions of \citet{leja2018}, the photometric data are fit using \texttt{Prospector-$\alpha$}. In particular, we adopt a non-parametric approach to model SFHs, imposing the continuity prior \citep{leja2019,leja2018}. The continuity prior disfavors unphysical jumps in SFH, i.e., episodes of rejuvenation and quenching, and it leads to a more physical and smoother SFH \citep{leja2019}. We refer the reader to \citet{leja2019} for further discussion of different priors of SFH, noting that the prior that we assume in our joint-fitting is similar to the continuity prior (see Equation \ref{eq:continuity-prior}).

\texttt{Prospector-$\alpha$} adopts the \citet{kriek2013} dust model, which is based on the parameterization of dust attenuation by \citet{noll2009}. In this model, the strength of the 2175$\mathrm{\AA}$ UV bump is correlated with the dust slope. Therefore, the free parameters are \texttt{dust\_index} (dust slope) as well as two dust attenuation parameters, \texttt{dust1} and \texttt{dust2} for the stellar populations younger and older than $10^7$ years, respectively. We note that \texttt{dust\_index} parameter controls the slope of dust attenuation curve, and for positive values the attenuation curve will be flatter than the Calzetti law \citep[e.g.,][Figure 1]{kriek2013}, leading to less UV attenuation and more near-IR attenuation comparably. For the negative values of \texttt{dust\_index} the opposite holds. 

We use the Mesa Isochrones and Stellar Tracks \citep[MIST;][]{choi2016} to generate the SSPs using Flexible Stellar Populations Synthesis models \citep[FSPS;][]{conroy2009,conroy2010}. The stellar metallicity is therefore measured relative to the solar abundance, as defined in Table~1 of \citet{choi2016}, and it is constrained by \texttt{Prospector-$\alpha$} based on the UV to optical to near-IR ratios of the SED \citep[see Figure 3 of][]{leja2017}.

\subsubsection{Priors to Generate SSPs for Spatially-resolved and Global Joint-fit \label{subsec:joint-fit-prior}}

We include all global photometric measurements (\emph{HST} and \emph{Spitzer}) in the \texttt{Prospector-$\alpha$} fit and use the resulting posterior as a prior in the joint-fitting.

For fitting the spatially resolved stellar populations, we calculate the spatially resolved photometric fluxes for the \emph{HST} bands by summing the flux for all pixels in each bin. To estimate the photometric uncertainty and to be sure that the correlated pixel noises are accounted for, we follow \citet{whitaker2011,skelton2014}, where the uncertainty is scaled by a power-law function of aperture sizes approximated by $\sqrt{N}$, where N is the number of pixels in each bin. We fit the resolved \emph{HST} photometry and error of each spatial bin using \texttt{Prospector-$\alpha$}. 

To take into account the dust and metallicity uncertainties in the spatially-resolved joint-fit, we have two choices of priors to generate SSPs, corresponding to two \texttt{Prospector-$\alpha$} fits. The first prior is defined using the spatially-resolved \texttt{Prospector-$\alpha$} fit, while the second prior is defined by the global \texttt{Prospector-$\alpha$} fit for all spatial bins (see section \ref{subsec:building-blocks} for the detail of including the dust and metallicity uncertainties for generating SSPs). The first prior is tuned to the resolved \emph{HST} bands of the spatial bins, but the corresponding \texttt{Prospector-$\alpha$} fit does not include IRAC channels. The second prior is not tuned to the individual bins, however it does include the IRAC channels 1 and 2. As there is no clear preference \emph{a priori}, we perform our spatially-resolved joint-fitting adopting both of these priors. All of the fits being performed, with their SSP priors and included data, are summarized in Table \ref{tab:fits}.

\begin{table*}[!t]
\begin{center}
\begin{tabular}{|c|c|c|}
 \hline
Weights & Predictors & Description \\
\hline\hline
 $\left[ x_{ij} \right]_{M\times N}$ &
 \begin{tabular}{@{}r@{}} $\left[ A_{s,ijl} \right] _{M\times N \times X}$ \\
  $\left[ A_{p,ijr} \right] _{M\times N \times P}$
 \end{tabular}
 &
 \begin{tabular}{@{}c@{}c@{}c@{}c@{}}$\mathbf{A_{s}}$ and $\mathbf{A_p}$ are SSP templates . \\ $M$ is the total number of spatial bins, \\ $N$ is the total number SSPs for each age, \\ $X$ is the total number of G141 pixels. \\ $P$ is the total number of photometric bandpasses. 
\end{tabular}
 \\ \hline
 $\left[x_{em,iq}\right]_{M\times4}$ &
 $\left[A_{em,iql}\right]_{M\times 4 \times X}$  &
 \begin{tabular}{@{}c@{}c@{}}
 The weight $\mathbf{x_{em}}$ of the emission line templates $\mathbf{A_{em}}$ \\ Four emission lines that are in WFC3/G141 bandpass for MRG-S0851, \\ are included. They are: $\left[ \mathrm{O III}\right]$, $\mathrm{H\beta}$, $\mathrm{H\gamma}$ and $\mathrm{H\delta}$
 \end{tabular} \\
 \hline $x_c$ &
 $\left[A_{c,l}\right]_{X}$ &
 The weight $x_c$ for the contamination $\mathbf{A_c}$ of MRG-S0851 by other objects.
 \\ \hline $\left[x_{b,k}\right]_{G}$
 & $\left[A_{b,kl}\right]_{G\times X}$ &
 \begin{tabular}{@{}c@{}c@{}}
 The different exposures could have different constant backgrounds. \\ The constants are $\mathbf{x_b}$ (``bias'' term in regression), \\ and the background model is $\mathbf{A_b}$. Here, G is the number of exposures. \end{tabular} \\ \hline
 $\left[ x_{p,in}\right]_{M\times I}$ &
 $\left[A_{p,inl}\right]_{M\times I \times X}$&
 \begin{tabular}{@{}c@{}}
 The weight $\mathbf{x_p}$ of the polynomial fit $\mathbf{A_p}$ to the data. \\ I is the degree of the polynomial used.
 \end{tabular} \\ \hline
\end{tabular}
\caption{Elements of the \texttt{requiem2d} generalized regression model used in the joint spectro-photometric fit. All matrices are denoted with the brackets and their shapes are shown as indices. Note that to reduce computation cost, we eventually turn these matrices to 1D arrays (see Figure \ref{fig:model-plate}). The first and second columns indicate the weight and its corresponding ``predictor'' followed by a description in the third column. \label{tab:regresssion}}
\end{center}
\end{table*}

\subsection{Elements of the \texttt{requiem2d} Full Model \label{subsec:model-elements}}

In this Section, we discuss the building blocks of our Bayesian model: the elements of the regression model, the prior, and the likelihood distributions.

\subsubsection{The Building Blocks of the Linear Regression Model \label{subsec:building-blocks}}

Modeling a composite stellar population using a linear combination of SSP templates is a generalized linear regression problem, whose elements are shown in Table \ref{tab:regresssion}. We use the FSPS models and its \texttt{python} wrapper \citep{mackey2015}, assuming the dust and metallicity values from the \texttt{Prospector-$\alpha$} posterior (Section \ref{subsec:dust-metal-prior}). The SSP spectra vary in age starting at 10 Myr to the age of the universe at the redshift of the main science target, increasing with a logarithmic scale, such that the logarithm of the ratio of two adjacent ages in Gyr is 0.05, i.e. $\log t_{i+1} [\mathrm{Gyr}]/t_i [\mathrm{Gyr}] = 0.05$.\footnote{We practically generate a series of non-overlapping constant SFHs that include each age in our grid at their center. This is argued to be a more realistic approximation than pure SSPs \citep[e.g., see][]{morishita2019}. We test both cases, but we do not find any significant difference, potentially owing to the finer sampling of the age grid in our study ($\sim 50$) in our case comparing to 10 of \citet{morishita2019}).}. These spectral models are then used to simulate the corresponding 2D G141 grism spectra using \texttt{Grizli}. We also add a first degree polynomial to the FSPS templates. By fitting a polynomial to grism spectra, we address any issues in the background such as enhanced airglow for a particular orientation and/or contamination \citep{brammer2012a}. In a joint spectro-photometric fit in particular, this polynomial fit addresses any spectroscopic flux calibration errors and tunes the spectral continuum shape to the photometry \citep{newman2018}, noting that the photometric data is solely modeled by SSPs with no extra polynomial being fitted.  This prevents the continuum shape of the grism spectrum from solely dictating the dust solution, as this should be mostly determined by the longer wavelength baseline of photometry. Also, by keeping the polynomial degree to lower values (usually less than 3, and for our pilot target, MRG-S0851, we pick 1), we prevent it from generating any spikes or spectral features that could affect age/metallicity measurements. 

\begin{figure}
    \centering
    \includegraphics[width=\columnwidth]{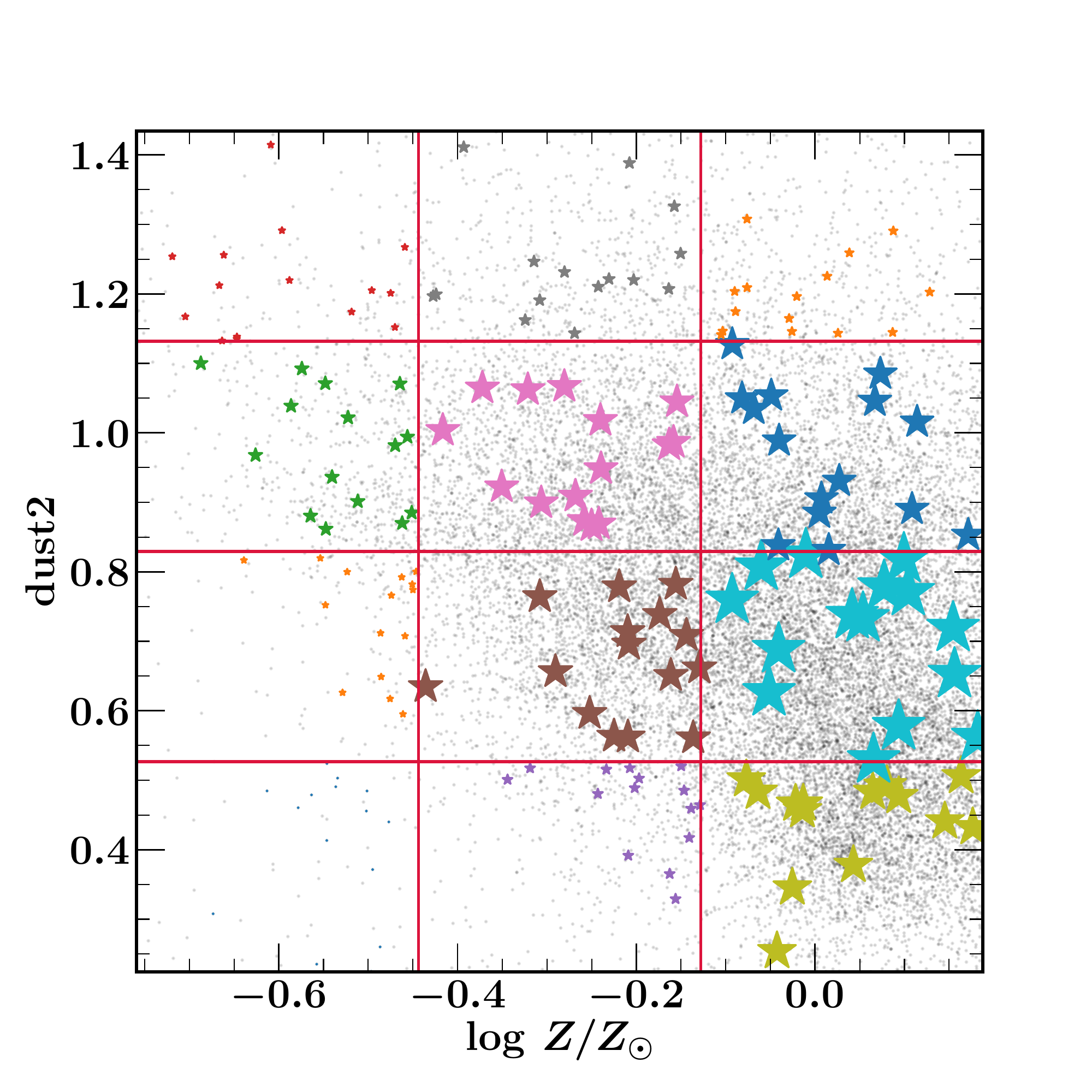}
    \caption{The \texttt{Prospector-$\alpha$} posterior of \texttt{dust2} and metallicity for MRG-S0851 are shown by light grey points. We show different boxes, defined to sample the posterior to generate SSPs, with red lines, and we indicate the actual draws in each box are by stars. Size of stars demonstrate the weight of each box in the \texttt{Prospector-$\alpha$} posterior.}
    \label{fig:dust2-metal-select}
\end{figure}

In order to use the \texttt{Prospector-$\alpha$} posteriors as the dust and metallicity priors to generate SSPs for resolved fitting with \texttt{requiem2d}, we project the full posterior into the \texttt{dust2} and $\log Z/Z_\odot$ plane, limiting the extension of each axis of this plane to the 3$\sigma$ width of the corresponding credible interval. We note that our \texttt{Prospector-$\alpha$} fit assumes the \citet{kriek2013} dust model which has 3 free parameters, including \texttt{dust2} that controls the attenuation of stellar populations older than $10^7\, \mathrm{yr}$ \citep[e.g., see][and Section \ref{subsec:prospector} for further discussion]{noll2009,conroy2010,kriek2013}. We then define $3\times 4$ boxes in this plane, drawing 15 samples from the full \texttt{Prospector-$\alpha$} posterior in each box, calculating the median of the draws. In other words, we use the 2D projection in  $\log Z/Z_\odot$-\texttt{dust2} plane to draw samples from the \emph{full} \texttt{Prospector-$\alpha$} posterior of dust and metallicity, that has 4 parameters. Figure~\ref{fig:dust2-metal-select} shows the 2D projection of the \texttt{Prospector-$\alpha$} posterior for the global analysis of the pilot target, MRG-S0851.

We have 12 sets of templates, each corresponding to a different region in dust and metallicity. Each one of the 12 sets of templates has a weight which is inferred by summing the weights of individual draws from the \texttt{Prospector-$\alpha$} posterior falling into the corresponding box posterior and the selections described here (Figure~\ref{fig:dust2-metal-select}). We rank order the SSPs using the final weights and sample the weights using a stick-breaking Dirichlet process \citep[e.g.,][]{connor1969,sethuraman1994} with $\beta \sim \mathrm{Beta}(1,\alpha)$ and $\alpha \sim \mathrm{Gamma}(11,1)$. In this process, one draws a set of initial 12 weights, $\beta^\prime _i,\, i=1\ldots 12$ from the Beta distribution. $\beta^\prime _i$s are between 0 and 1, but they do not necessarily add up to one, and to make sure that they do, the final set of weights is calculated using $\beta_i = \beta^\prime_i \Pi _{j=1}^{i-1}(1-\beta^\prime _j)$, analogous to breaking a stick of a length 1.

For our pilot target, MRG-S0851, we have four major emission lines filling the underlying absorption features in G141 bandpass:  $\mathrm{H\beta}$, $\mathrm{H\gamma}$, $\mathrm{H\delta}$, and $\mathrm{[O\, III]}$. We include a separate template for each one of these emission lines from \texttt{Grizli}: A Gaussian one dimensional spectral template centered at the wavelength of each emission line is normalized to one and is convolved with MRG-S0851 morphology to generate a two dimensional grism template. The coefficients of these templates are being fitted with the rest of parameters using the Monte Carlo method, providing an estimate on the strength of emission lines.

We multiply all photometric bands in the model of each spatial bin with a set of nuisance parameters $\mathbf{\omega}$, with a prior of $\mathrm{N}(1,1)$, i.e., a normal distribution with $\mu=1$ and $\sigma=1$, to address any calibration mismatch between the photometric and spectroscopic data. The general model for one set of SSPs with all elements in place can then be described by the following equations (see Table \ref{tab:regresssion} for the description of each element):
\begin{align}
    M_{s,l}&=\sum_{i=1}^{M} \sum _{n=1}^{I} x_{p,in}A_{p,inl} +\sum _{i=1}^{M} \sum _{j=1}^{N} x_{ij} A_{s,ijl}  \nonumber \\ &+ \sum _{i=1}^{M}\sum _{q=1}^{4} x_{em,iq} A_{em,iq}+x_c A_{c,l}
    \nonumber \\
    &+\sum _{k=1}^{G}x_{b,k}A_{b,kl} \label{eq:model-spec}\\
    M_{p,ir}&=\omega _i\sum_{j=1}^{N}x_{ij}A_{p,ijr},
    \label{eq:model-phot}
\end{align}
where $M_{s,l}$ denotes the 2D grism model of the $l$-th \emph{HST} pixel, and $M_{p,ir}$ indicates the photometric model for the $i$-th spatial bin and the $r$-th photometric band.

\begin{figure*}
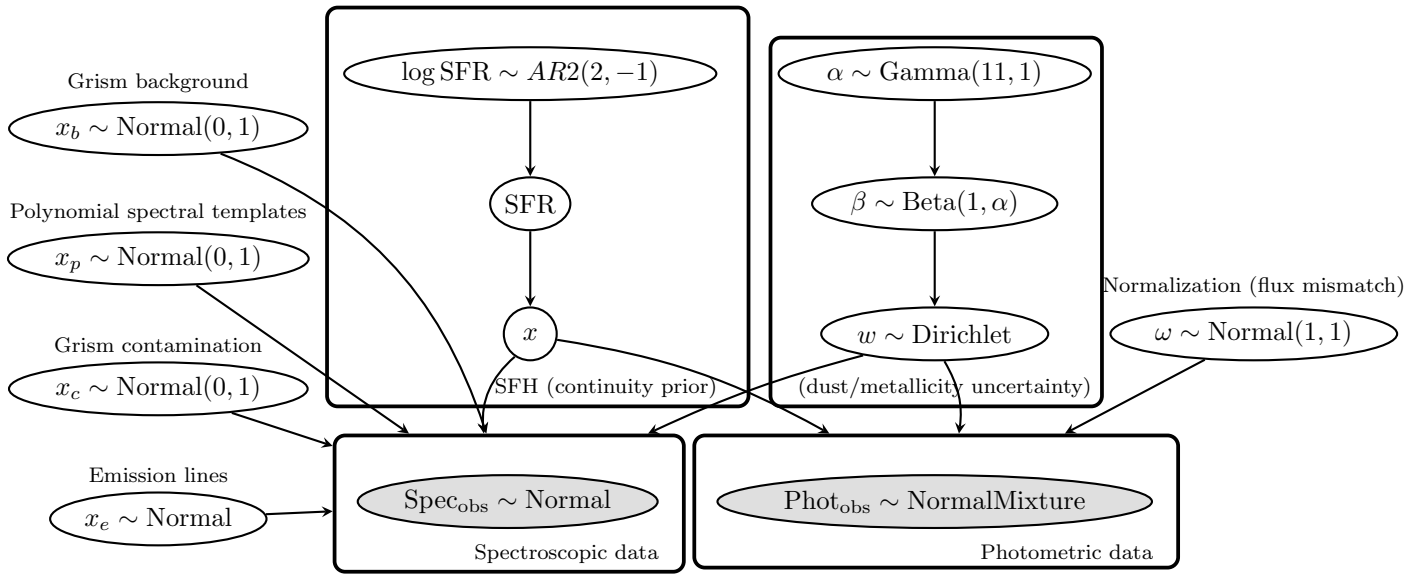

\centering
  \tikz[->,>=stealth,thick]{
    
    \node[latent,ellipse] (lsfr) {$\log \mathrm{SFR}\sim AR2(2,-1)$}; %
    \node[latent,ellipse,below=of lsfr] (sfr) {$\mathrm{SFR}$}; %
    \node[latent,ellipse,below=of sfr] (x) {$x$};
    
    \node[latent,ellipse,xshift=-0.2cm,right=of lsfr] (alpha) {$\alpha \sim \mathrm{Gamma}(11,1)$}; %
    \node[latent,ellipse,below=of alpha] (beta) {$\beta \sim \mathrm{Beta}(1,\alpha)$}; %
    \node[latent,ellipse,below=of beta] (w) {$w\sim \mathrm{Dirichlet}$}; %

    \node[obs,ellipse,below=of w,yshift=-0.5cm] (phot_obs) {$\mathrm{Phot_{obs}}\sim \mathrm{NormalMixture}$};%
    \node[obs,ellipse,xshift=-0.3cm,yshift=-0.5cm,below=of x] (spec_obs) {$\mathrm{Spec_{obs}}\sim \mathrm{Normal}$};%
    \node[latent,ellipse,label=Normalization (flux mismatch),,xshift=-0.2cm,right=of w] (omega) {$\omega \sim \mathrm{Normal}(1,1)$}; %

    \node[latent,ellipse,xshift=-1.4cm,yshift=1.0cm,label=Grism background, left=of sfr] (x_b) {$x_b\sim \mathrm{Normal(0,1)}$};%
    
    \node[latent,ellipse,label=Polynomial spectral templates,below=of x_b] (x_p) {$x_p\sim \mathrm{Normal(0,1)}$};%

    \node[latent,ellipse,label=Grism contamination,below=of x_p] (x_c) {$x_c\sim \mathrm{Normal(0,1)}$};%
    
    \node[latent,ellipse,label=Emission lines, below=of x_c] (x_e) {$x_e\sim \mathrm{Normal}$};%

    \plate [inner sep=0.3cm,xshift=.1cm,yshift=.2cm,line width=0.5mm] {plate3} {(x) (sfr) (lsfr)} {SFH (continuity prior)}; %
    
     \plate [inner sep=0.3cm,xshift=.02cm,yshift=.2cm,line width=0.5mm] {plate1} {(spec_obs)} {Spectroscopic data}; %
     
     \plate [inner sep=0.3cm,xshift=.02cm,yshift=.2cm,line width=0.5mm] {plate2} {(phot_obs)} {Photometric data}; %
     
     \plate [inner sep=0.1cm,xshift=.02cm,yshift=0.0cm,line width=0.5mm] {plate4} {(w) (beta) (alpha)} {(dust/metallicity uncertainty)}; %

     \path[every node/.style={font=\sffamily\small}]
    (lsfr) edge node [left] {} (sfr)
    (sfr) edge node [right] {} (x)
    (x) edge[bend left=10] node [right] {} (plate2)
    (x) edge[bend right=27] node [right] {} (plate1)
    (w) edge[bend right=5] node [right] {} (plate1)
    (w) edge[bend left=19] node [right] {} (plate2)
    (x_b) edge[bend left=25] node [right] {} (plate1)
    (x_p) edge node [right] {} (plate1)
    (x_c) edge node [right] {} (plate1)
    (x_e) edge node [right] {} (plate1)
    (omega) edge node [right] {} (plate2)
    (alpha) edge node [right] {} (beta)
    (beta) edge node [right] {} (w)
    
    ;
     }
\caption{The statistical model for the spatially-resolved analysis demonstrated using the plate notation. \label{fig:model-plate}}
\end{figure*}
\subsection{The Priors and the Monte Carlo Sampling Method  in \texttt{requiem2d} \label{subsec:priors}}

We determine the posterior distribution of weights in the generalized linear regression model (defined in Equations \ref{eq:model-spec} and \ref{eq:model-phot}) using  Bayes' theorem. The photometry likelihoods are assumed to follow a mixture of normal distributions with a standard deviation estimated from the observational errors and weights from the Dirichlet process. To be more specific, for each one of 12 set of SSPs (Section \ref{subsec:building-blocks}), we generate a model using the SFH model and calculate the photometric fluxes. We next assume 12 normal distributions centered at these fluxes with standard deviations equal to the observed photometric uncertainty. The full likelihood probability distribution is then the weighted sum of the 12 normal distributions with weights determined through a stick-breaking Dirichlet process, as described in Section \ref{subsec:building-blocks}.  

As we have more than 1000 grism pixels for each spatial bin in spatially-resolved spectroscopic data, we adopt a simplifying assumption that the final grism model is a weighted average of 12 SSPs. To maintain consistency between the resolved and unresolved analysis, we apply the same assumption to the spatially-unresolved spectroscopic data. We test this simplifying assumption explicitly for the unresolved analysis of MRG-S0851 by sampling the age posterior twice, first using a full mixture of normal distributions for grism spectroscopy and then using the weighted average of 12 SSPs. No statistically significant difference is detected in recovered ages adopting these two approaches.

The prior of weights, $\mathbf{x}$, is derived from the SFH prior. As mentioned in Section \ref{subsec:prospector-phot-fit}, we adopt a continuity prior for the SFH following a regularizing scheme introduced for the same problem in \citep{ocvirk2006}:

\begin{equation}
    \log\mathrm{SFR}_{n,t} - 2\log \mathrm{SFR}_{n,t-1}+\log \mathrm{SFR}_{n,t-2} \sim \epsilon_t, \label{eq:continuity-prior}
\end{equation}
with $\epsilon _t \sim \mathrm{N}(0,1/20)$. For a linearly defined age grid, this can also be interpreted physically as a requirement of the continuity for the first time derivative of SFR (the slope of SFR, or SFR increments). Other versions of a continuity prior may also be used. For example \citet{leja2019} require the continuity of the SFR itself\footnote{We note that \citet{leja2019} adopt Student's t-distribution for $\epsilon_t$ in the right hand side of Equation \ref{eq:continuity-prior}.}. We find that in our case, analyzing massive quiescent galaxies at z$\sim$2, the continuity of the SFR slope recovers SFHs and ages slightly better than the continuity of the SFR. This may be because it is a stronger prior requirement, helping with the finer sampling of SSPs at lookback times greater than $\sim$1 Gyr.

To connect SFRs to the weights, each SSP has a mass-to-light ratio which we use to calculate the corresponding mass weight, $\mathbf{x^M}$, from $\mathbf{x}$. This mass weight can then be connected to SFR using $\mathrm{SFR}_t=x^M_t/\delta t$. The rest of priors and the model itself are demonstrated in Figure \ref{fig:model-plate}. 

Estimating the age, weights $\mathbf{x}$, or other parameters of the stellar populations from the observed spectra could be ill-posed and it usually requires regularizing \citep{ocvirk2006}. Also, due to highly correlated parameters and the higher dimension of the problem, the usual Monte Carlo algorithms such as random-walk Metropolis \citep{metropolis1953} fail to sample the posterior efficiently \citep{neal1993}. We therefore use No-U-Turn sampling \citep[NUTS;][]{homan2014}, which is a variation of the Hamiltonian Monte Carlo method \citep{neal2011} to sample the posterior. NUTS uses a recursive algorithm to build a set of points spanning a wide swath of the target distribution, stopping automatically when it starts to retrace its step \citep{homan2014}. NUTS is proven to be more efficient in exploring correlated parameter spaces such as in our problem relative to the random-walk methods \citep{cruetz1988,neal2011,homan2014}.

We use the \texttt{python} package \citep[\texttt{pymc3};][]{salvatier2016} extensively in our analyses for sampling the posteriors with NUTS. Two chains are constructed, drawing 1000 (unresolved analysis) and 1400 (resolved analysis) samples in each one, considering only the second half of the chains as post burn-in draws. We check for divergences using Gelman-Rubin statistics \citep{gelman1992} explicitly and combine the chains.

\section{Inferring the Age of Stellar Populations \label{sec:age-appendix}}

In this section, we show how we can infer the age of stellar populations and its uncertainty from the weights, $\mathbf{x}$, defined in Equations \ref{eq:model-spec} and \ref{eq:model-phot}.

\begin{figure*}
\centering
\includegraphics[width=0.75\textwidth]{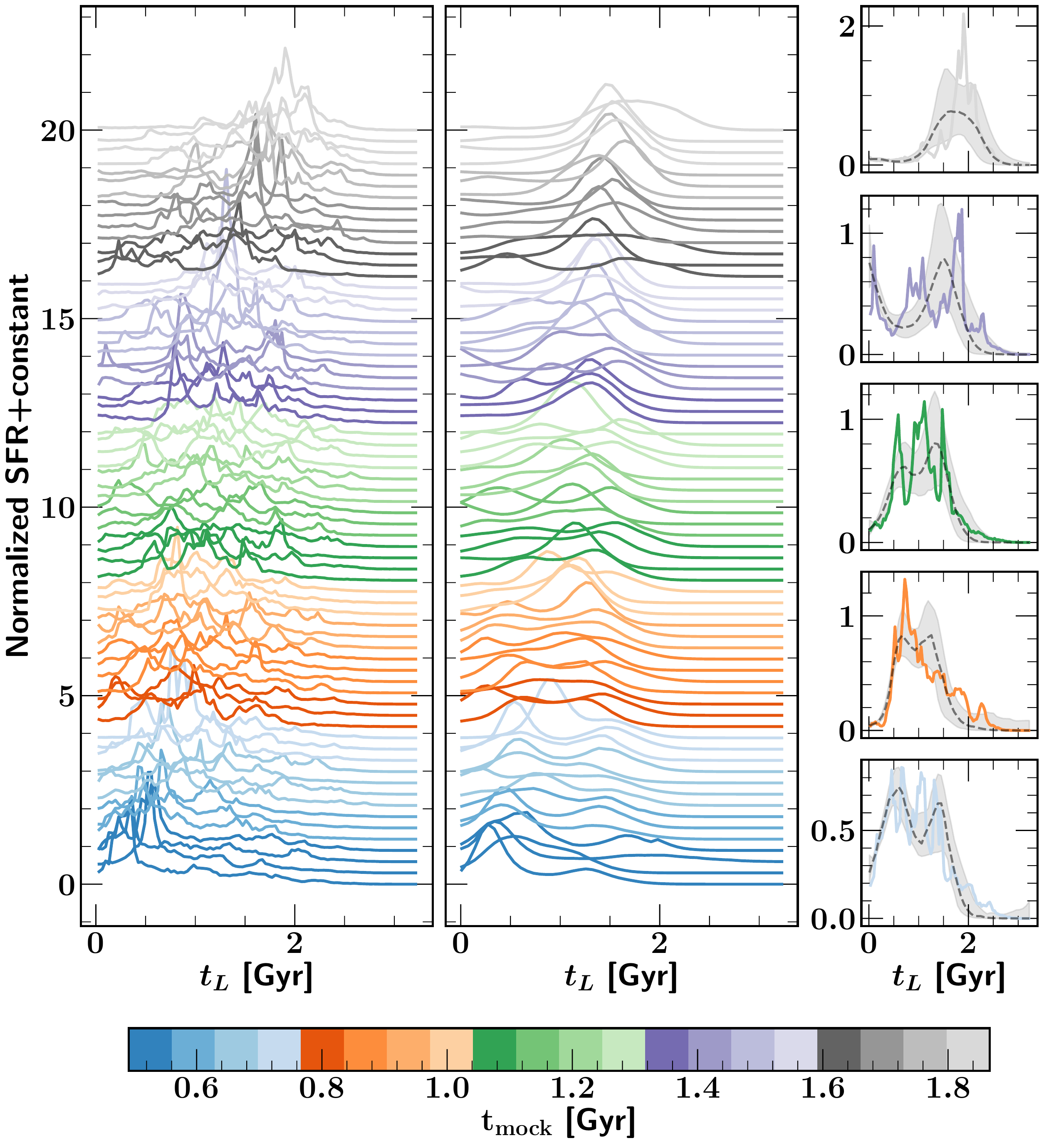}
\caption{Testing the \texttt{requiem2d} methodology using SFHs of massive quiescent and transition galaxies at $z=2$ selected from the Illustris simulation. This figure shows Illustris SFHs and recovered SFHs using \texttt{requiem2d} versus lookback time, $t_L$. SFHs are color coded by their mock Illustris median ages, $t_{\mathrm{mock},50}$. The Illustris SFHs (left panel) agree well with the recovered SFHs (middle panel). The right panel shows 5 SFHs with a range of ages and their recovered models in more detail. The dashed grey line and grey shade indicate the median and 1$\sigma$ width of recovered SFH using \texttt{requiem2d}. The second galaxy from the top in the right panel has SFH that looks similar to what we recover for MRG-S0851 in the last $\sim$1 Gyr of evolution, and the top right panel demonstrates the oldest galaxy in the sample with a median age of 1.9 Gyr.} \label{fig:mock_test1}
\end{figure*}

The posterior of weights, $\mathbf{x}$, which is the basic output of \texttt{requiem2d}, can be interpreted as the light-weight of each SSP template. We can use these weights to calculate the light-weighted average ages, however, light-weighted ages are misleadingly young as younger stars outshine the older stars. We therefore use the mass-to-light ratio of SSP templates to calculate the mass-weights, $\mathbf{x}^M$. This quantity is used to reconstruct SFHs and to calculate the median mass-weighted age $t_{50}$\footnote{$t_{50}$ is formally defined as $\int_{t_{50}}^{t_0}dt^\prime \, \mathrm{SFR}(t^\prime)=0.5\times \int_0^{t_0}dt^\prime\, \mathrm{SFR}(t^\prime)$, where $t_0$ is the age of the universe at the redshift of interest.}, shown to be a robustly estimated from models \citep[e.g.,][]{belli2019}. The uncertainty of the median mass-weighted age is estimated directly from the Monte Carlo chains. The median mass-weighted age, $t_{50}$, is also independent of the lensing magnification, as any effect of magnification on SFR is cancelled out in $t_{50}$ definition (like sSFR).

The final goal of the \texttt{requiem2d} code is to recover both global and resolved ages and SFHs of massive quiescent galaxies, and as it uses a non-parametric SFH history in a Bayesian framework, it is important to choose an appropriate SFH prior. We test our methodology and our choice of prior, Equation \ref{eq:continuity-prior}, using a sample of massive galaxies with $\log M_*/M_\odot \geq 10.6$ at $z=2$, selected from Illustris, a cosmological hydrodynamical simulation of galaxy formation with a volume of $(100 \mathrm{Mpc})^3$ that includes a comprehensive physical model \citep[][]{vogelsberger2014,vogelsberger2014a,genel2014}. To obtain the Illustris SFHs for our test, we construct the histograms of the formation times of all the star particles in the galaxy, weighted by the masses of the star particles at $z$=2. The Illustris SFHs should therefore be directly comparable to our mass-weighted SFHs. Using Illustris SFHs, we perform three tests: we test if we can recover (1) global SFHs and ages of quiescent and transition galaxies, (2) age gradients of quiescent galaxies, and (3) global ages and SFHs of star-forming galaxies. We discuss these tests in turn.

\begin{figure}
    \centering
    \includegraphics[width=1.0\columnwidth]{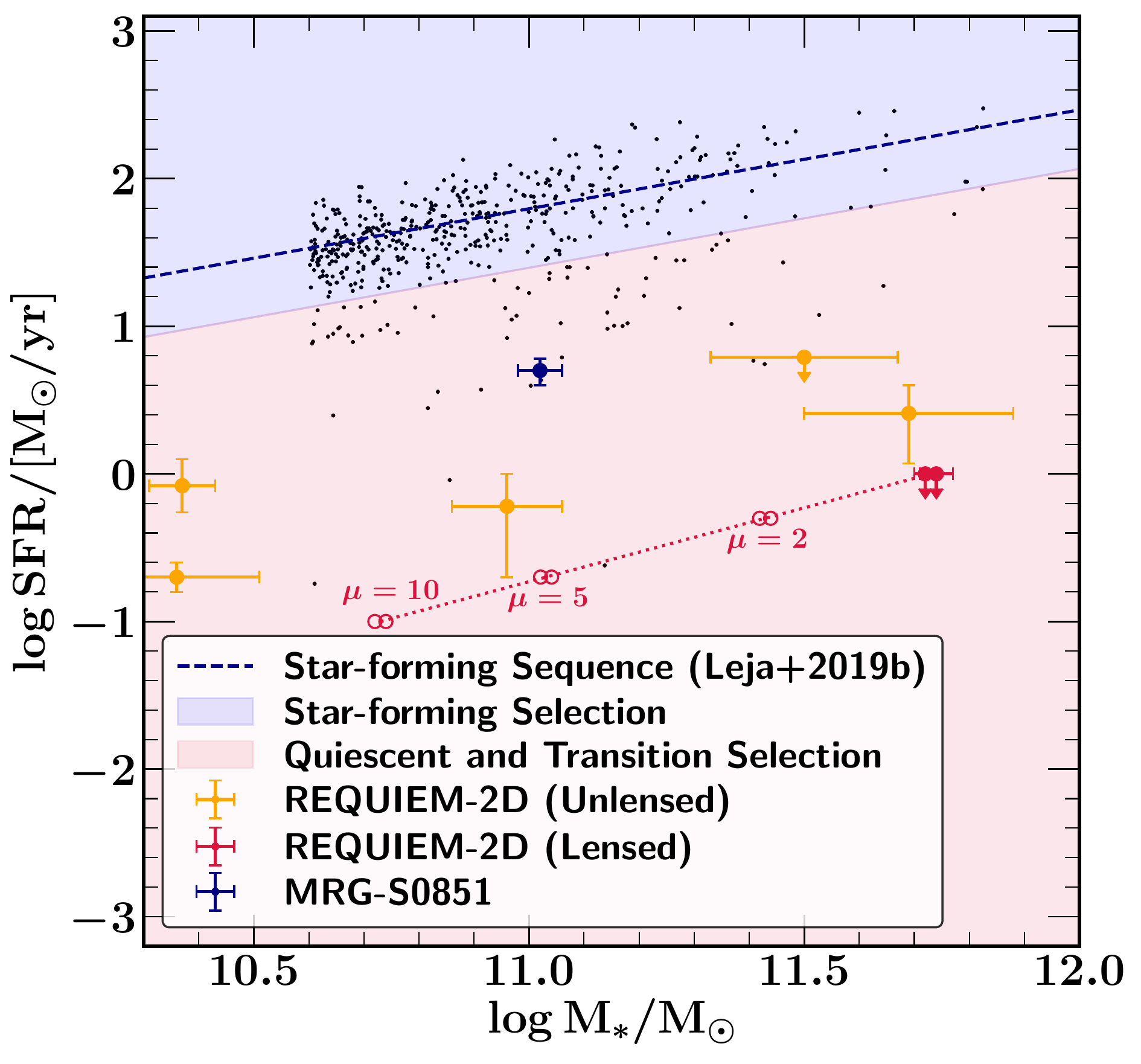}
    \caption{A sample of massive galaxies in the log(SFR)-log(stellar mass) plane, selected from the Illustris cosmological simulation to test our methodology and the choice of SFH prior. The black dots are the Illustris galaxies, with our REQUIEM sample indicated by larger circles using global SFRs measured by \citet{newman2018}, A. Man et al. (in prep) and this work. The SFR and stellar mass of two targets, MRG-S1522 and MRG-P0918, are not demagnified; we plot their projected locations assuming magnifications of $\mu$=2, 5, 10 (open circles) along the dotted line trajectory of increasing magnification. The color-coding identifies our sub-samples of quiescent and transition (red) and star-forming populations (blue).}
    \label{fig:illustris-sel}
\end{figure}

 A sub-sample of massive galaxies with low sSFRs is selected by requiring that the SFR at the closest snapshot at $z=2$, corresponding to the average SFR in the last $\sim$30Myr given the width of our time bins, is 0.4 dex below the empirical star-forming sequence in the log(SFR)-log(stellar mass) plane  of \citet{leja2018}, noting that the locus of the star-forming main sequence in \citet{leja2018} is 0.3 dex lower than \citet{whitaker2014}. Our Illustris selection consists of a more diverse sample than traditional quiescent galaxies, including galaxies below the main sequence but above the quiescent population according to \citet{leja2018}. Therefore, we call this population quiescent and transition galaxies hereafter.  This selection leaves room for new discoveries and makes our test more robust. The region in log(SFR)-log(stellar mass) parameter space corresponding to our combined quiescent and transition selection is shown in Figure~\ref{fig:illustris-sel}.  Out of 502 galaxies initially selected from Illustris given our stellar mass cut of $\log M_*/M_\odot \geq 10.6$ at $z=2$, 71 galaxies are included in our quiescent and transition sub-sample. The stellar-masses of the sub-sample have a distribution of $\log M_*/M_\odot=11.03^{+0.30}_{-0.37}$, where the upper and lower limits correspond to 84th to 50th and 50th to 16th percentiles. Figure~\ref{fig:mock_test1} (left panel) shows the final SFHs of quiescent galaxies selected from Illustris, rank ordered by their median ages. 

We generate spectral templates using FSPS models from the Illustris SFHs, fixing the dust and metallicity to the best fit values of MRG-S0851 (to be presented in Section \ref{subsec:prospector}). The morphology of MRG-S0851 is used to generate a mock grism spectrum, and we choose the same 5 \emph{HST} and 2 \emph{Spitzer} photometric bands of  to generate simulated photometric data. Noise is added to grism pixels by assuming a SNR that is drawn randomly for each target from a uniform distribution of  $\mathrm{U}(0.01,0.1)$ for grism pixels. We assume that the SNR of photometric measurements are 5 times higher than the SNR of grism pixels, in the range of 50-500. For comparison, the SNR of 12-orbit depth grism data of MRG-S0851 is $\sim$10-20 (corresponding to noise at the 5\%-10\% level) and the photometric SNR is $\sim$20-200 with $<$1 orbit of imaging data. 

\begin{figure*}
\centering
\includegraphics[width=\textwidth]{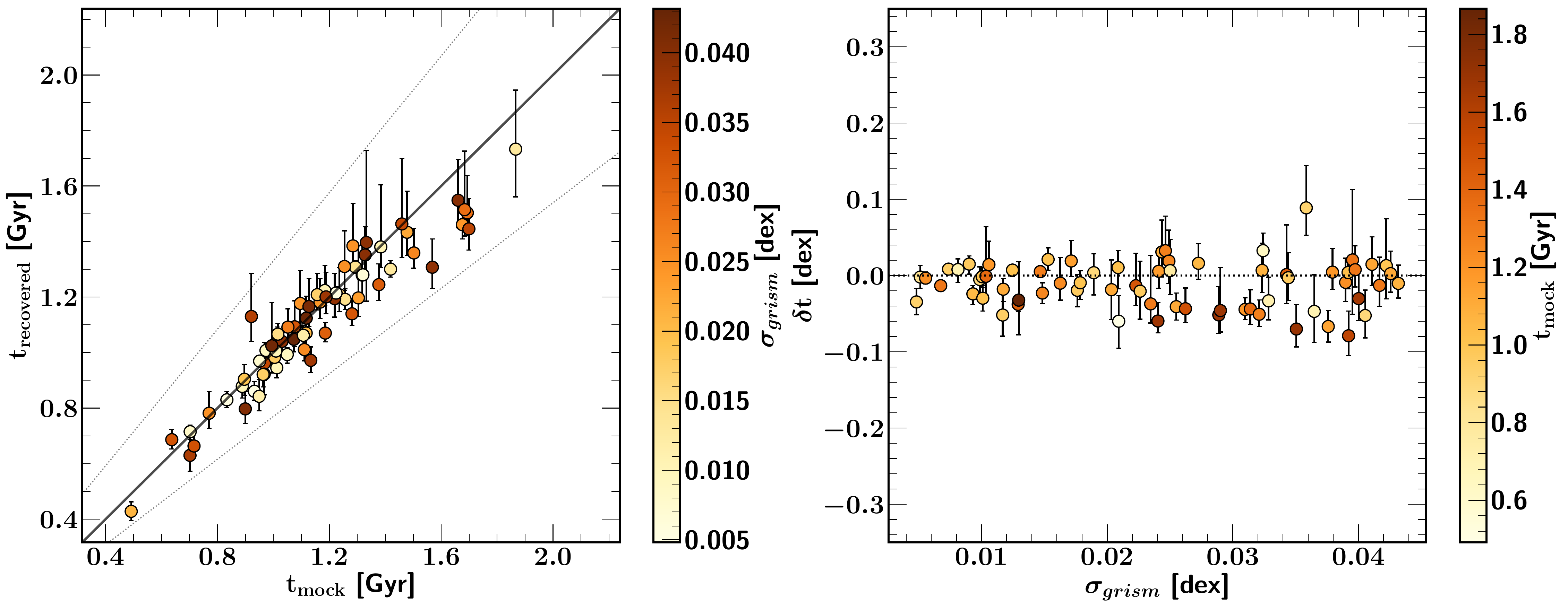}
\caption{Testing our methodology to recover global SFHs and ages using a sample of massive quiescent and transition galaxies selected from the Illustris simulation. \textbf{Left panel:} Recovered versus the actual global $t_{50}$ ages, color-coded by grism noise. The solid line is one-to-one relation and the dotted lines are $\pm$0.1 dex scatter. \textbf{Right panel:} Deviation of median recovered $t_{50}$ ages from the true ages versus grism noise, color-coded by true ages. No noticeable systematic biases can be seen, indicating that our choice of prior for SFHs (Equation \ref{eq:continuity-prior}) reasonably recovers the ages of massive quiescent and transition galaxies at $z=2$, selected from the Illustris simulation.} \label{fig:mock_test2}
\end{figure*}

The results of the simulations are presented in Figures~\ref{fig:mock_test1} and \ref{fig:mock_test2}. Figure \ref{fig:mock_test1} demonstrates that the general average trends of SFHs are recovered reasonably well.  However, one shortcoming of this methodology is that it struggles to recover particularly stochastic SFHs (see the right column, Figure~\ref{fig:mock_test1}). This is potentially due to the continuity prior that disfavors stochastic jumps.  Figure~\ref{fig:mock_test2} shows recovered median ages versus mock ages from Illustris (left panel) and its scatter as a function of grism noise per pixel. There are no noticeable systematic biases in the recovered ages, except for ages older than $\sim$1.5 Gyr that appear to be slightly younger (Figure \ref{fig:mock_test2}). We note that having a noisier grism spectrum increases the scatter around the recovered age, and by increasing the level of uncertainty from 0.01 dex to 0.05 dex per grism pixel, the scatter of median ages increases from $\sim$0.02 to $\sim$0.1 dex (Figure \ref{fig:mock_test2}, right panel). The results of these mock tests are therefore useful for planning future grism observations, where Figure \ref{fig:mock_test2} gives the required SNR for a given target age accuracy.

\begin{figure}
\centering
\includegraphics[width=\columnwidth]{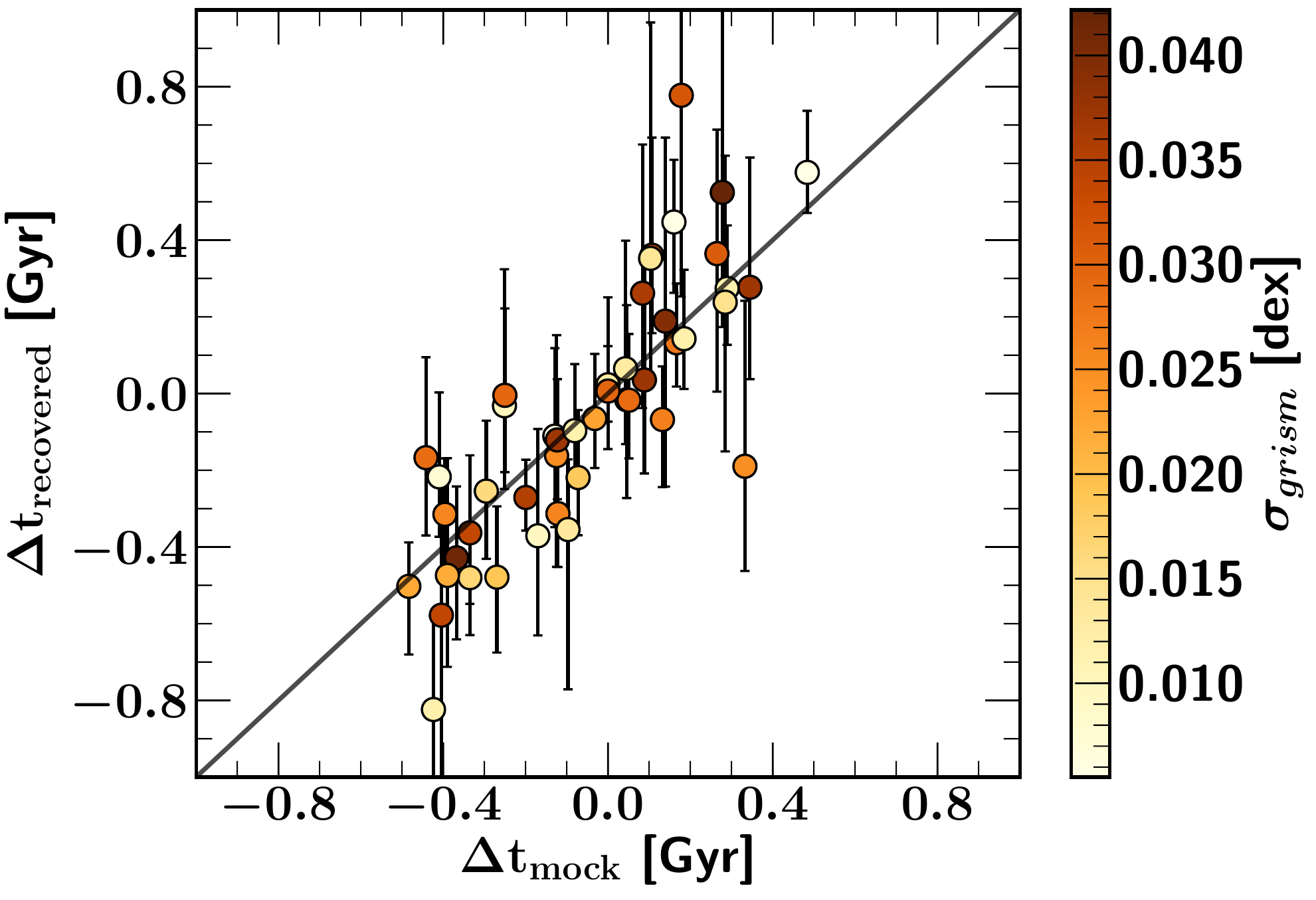}
\caption{Recovering age gradients of quiescent and transition galaxies using our methodology for three spatial bins with widths of $0\farcs18$ that have 5 resolved \emph{HST} photometric measurements, two unresolved \emph{Spitzer} IRAC channels 1 and 2 measurements, and grism specrtum. SFHs are selected from the Illustris simulation.}\label{fig:mock_test3}
\end{figure}

Next, we test the age gradients. To generate the mock observations, we use the MRG-S0851 morphology, defining three spatial bins with widths of $0\farcs18$ at the center of MRG-S0851. We assume that two adjacent bins have the same SFHs and ages (perfect symmetrical galaxy). We randomly select a SFH from the sub-sample of Illustris quiescent and transition SFHs, selected as described in the first test, for the central bin.  For adjacent bins, we then randomly select another Illustris quiescent and transition SFH such that the age gradient between the center and the two adjacent bins lies within $\pm 0.5$Gyr. Next, we generate FSPS models as described in the first test. Finally, we use \texttt{Grizli} to simulate the grism spectrum of three bins. To be fully consistent with the real data, 5 resolved \emph{HST} photometric bands for each one of the three bins are calculated but only global/unresolved \emph{Spitzer} IRAC channels 1 and 2 photometric measurements are assumed. We generate 50 mock observations and add noise in the exact same way as the first test by assuming a constant SNR per grism pixel. However, we note that in this test, we are only covering a portion of the galaxy, and the total flux associated with each SFH is relatively lower than the first test. 

Figure~\ref{fig:mock_test3} shows the age gradient test results, where we demonstrate the recovered $\Delta t = t_{\mathrm{50, center}}-t_{\mathrm{50, outskirt}}$. There are no systematic biases in the recovered age gradients, however, we caution that having a sharper gradient seems to increase the scatter.

As for the third test, we randomly select 100 galaxies from the complement of the quiescent and transition region in the log(SFR)-log(stellar mass) plane, i.e. we select 100 galaxies whose SFR is greater than 0.4 dex below the empirical star-forming sequence of \citet{leja2018} (blue region in Figure~\ref{fig:illustris-sel}). The mock data is generated following the exact same steps of the first two tests.

\begin{figure*}
\centering
\includegraphics[width=0.75\textwidth]{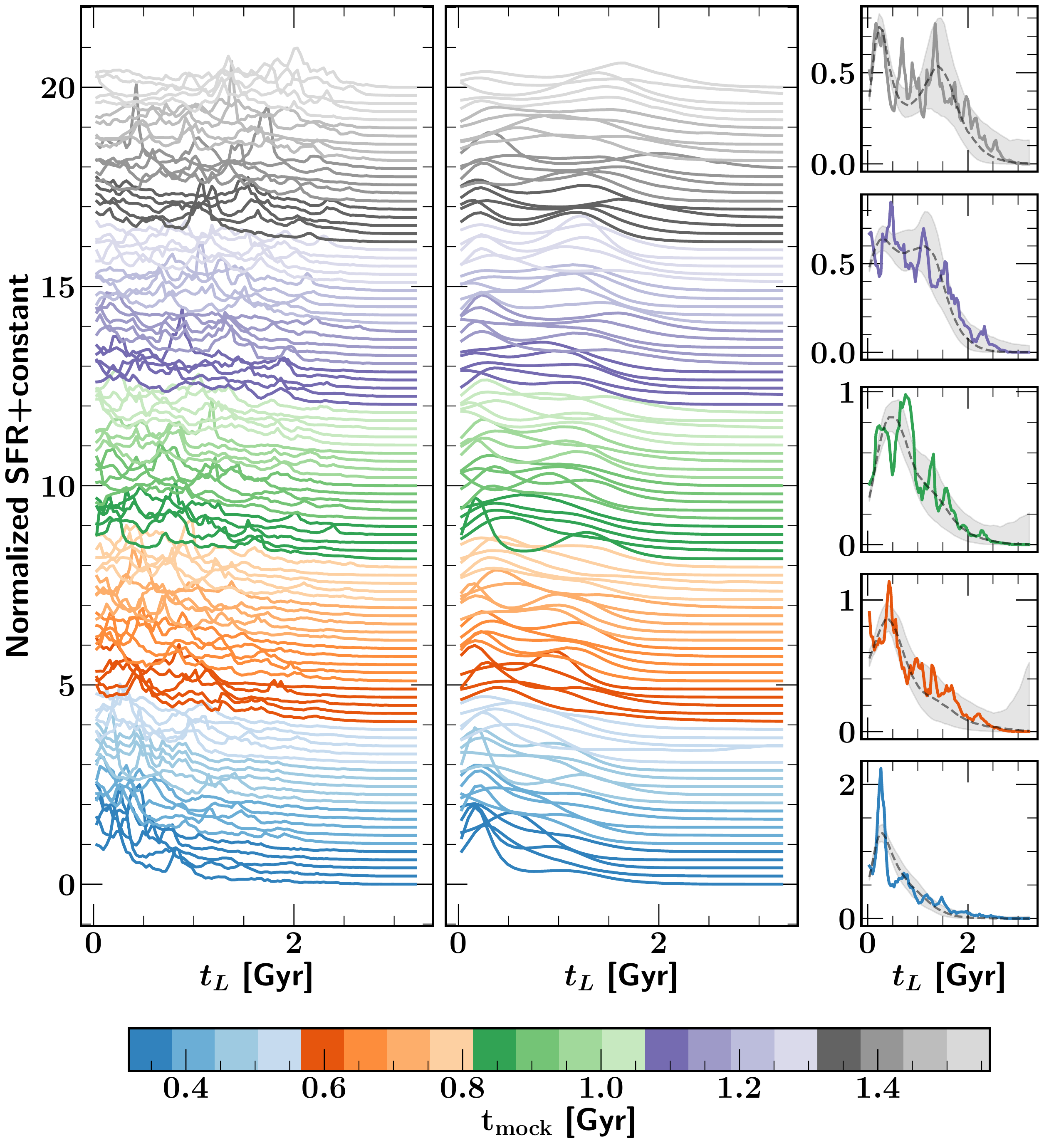}
\caption{Testing the \texttt{requiem2d} methodology using SFHs of massive star-forming galaxies at $z=2$ selected from the Illustris simulation. The panels and the labels are exactly the same as Figure~\ref{fig:mock_test1} but for massive star-forming galaxies.} \label{fig:mock_test4}
\end{figure*}

\begin{figure*}
\centering
\includegraphics[width=\textwidth]{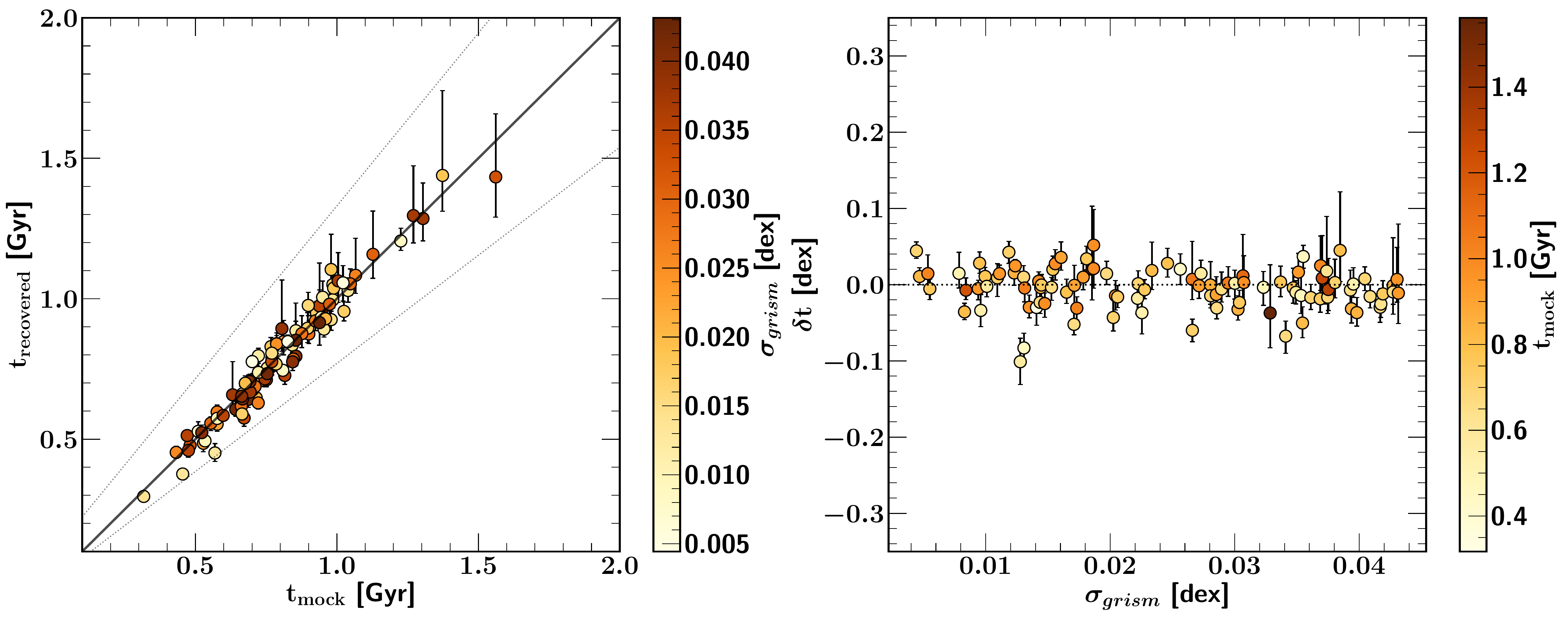}
\caption{Testing our methodology to recover global SFHs and ages using a sample of massive star-forming galaxies selected from the Illustris simulation. \textbf{Left panel:} Recovered versus the actual global $t_{50}$ ages, color-coded by grism noise. \textbf{Right panel:} Deviation of median recovered $t_{50}$ ages from the true ages versus grism noise, color-coded by true ages. The symbols and the labels are exactly the same as Figure~\ref{fig:mock_test2} but for massive star-forming galaxies.}\label{fig:mock_test5}
\end{figure*}

Figure~\ref{fig:mock_test4} shows an aggregate plot of the SFHs of massive star-forming galaxies at $z\sim2$ selected from the Illustris simulation. The \texttt{requiem2d} code and the continuity prior can recover the general trends of SFHs reasonably well with a notable exception of some difficulty recovering highly stochastic features in the SFHs, similar to the analyses of massive quiescent galaxies (see Figure~\ref{fig:mock_test1}).

Figure~\ref{fig:mock_test5} shows the recovered median mass-weighted ages versus median mock ages from massive star-forming galaxies in the Illustris simulation (left panel), and the scatter of median ages versus the grism pixel noises (right panel). The recovered ages match reasonably well with the mock ages. We also notice no systematic offsets or biases when considering the recovered ages in Figure~\ref{fig:mock_test5}.  These results demonstrate that the \texttt{requiem2d} code, with the continuity prior (Equation \ref{eq:continuity-prior}), can also be used to derive robust constraints on the ages of massive star-forming galaxies at $z\sim2$.

To conclude, we caution that a careful choice of SFH prior is important in non-parametric SFH models \citep[e.g.,][]{leja2019}. Our choice of the continuity of the SFR slope, Equation \ref{eq:continuity-prior}, seems to be appropriate for analyzing the SFHs and ages of massive galaxies at $z\sim2$, as justified by recovering ages and SFHs from the Illustris cosmological simulation.

\section{Testing our Methodology with Real Observations: A Pilot Study of MRG-S0851 \label{subsec:real-test}}

SDSS~J0851+3331-E was first detected in the Sloan Giant Arcs Survey (SGAS), which is a survey of strongly lensed galaxies \citep{hennawi2008, bayliss2011, sharon2020}. The cluster is at a redshift of $0.3689\pm0.0007$ with a right ascension of 8:51:39 and a declination of +33:31:10.83 \citep[see Tables 1-3 and Section 4.3.6 in,][for more detailed information and discussion]{sharon2020}.
Throughout this paper, our target is referred to as MRG-S0851, as it is a Magnified Red Galaxy \citep[MRG, following][]{newman2018} and represents a pilot analysis for the REQUIEM survey.

MRG-S0851 was observed with \emph{HST}, in program HST-GO-13003, PI: M. Gladders \citep{sharon2020}, as well as the \emph{Spitzer Space Telescope} \citep{rigby2012}. The \emph{Spitzer} data for MRG-S0851 includes observations with the IRAC channels 1 and 2 on December 13, 2010, as a part of a larger campaign of infrared imaging for the SGAS survey (\emph{Spitzer} proposal IDs 90232, PI: J. Rigby, and 70154, PI: M. Gladders). The \emph{HST}/WFC3 observations include the IR filters $\mathrm{H}_{\mathrm{F160W}}$ and $\mathrm{J}_{\mathrm{F125W}}$, as well as the UVIS filters $\mathrm{I}_{\mathrm{F814W}}$ and $\mathrm{U}_{\mathrm{F390W}}$ on February 26, 2016. 

We combine the existing observational data with our \emph{HST} program for MRG-S0851 (HST-GO-14622, PI: K. Whitaker), adding the WFC3 IR filters $\mathrm{H}_{\mathrm{F160W}}$, $\mathrm{Y}_{\mathrm{F105W}}$ and 12 orbits with the G141 grism dispersing element. For this follow-up program, each of the 12 orbits contained a short imaging exposure with either $\mathrm{H}_{\mathrm{F160W}}$ or $\mathrm{Y}_{\mathrm{F105W}}$ ($\simeq 250$ seconds of exposure time) followed by a longer exposure with the grism $\mathrm{G141}$ ($\simeq 2400$ seconds of exposure time). These 12 orbits of data were executed between January 28, 2017 to February 15, 2017. To minimize the contamination of the MRG-S0851 grism spectra from nearby bright cluster members, we used two different dispersion angles of $P_\theta \simeq 29^\circ$ and $P_{\theta}\simeq 35^\circ$ in the $\mathrm{G141}$ observations.

Figure \ref{fig:0851-E-mosaics} shows the final data product for MRG-S0851 for $\mathrm{H}_{\mathrm{F160W}}$ and the G141 grism. In the left panel, we show the $\mathrm{H}_{\mathrm{F160W}}$ mosaic, the deepest photometric data. In the right panel, we show the drizzled image of WFC3/G141 grism spectra for 6 orbits, corresponding to $P_\theta \simeq 29^\circ$ orientation angles. The 2D grism spectrum of the brightest image, E3, is shown in Figure \ref{fig:2d-grism}. 

The detail of the lensing model and the morphological analysis of MRG-S0851 are discussed in Appendix \ref{sec:cluster-lens} and \ref{sec:whole arc}, respectively. While MRG-S0851 is quintuply imaged, we will concentrate on image E3 in this paper, since other images are only partial according to our lens model, and their light profiles do not represent the full light profile of MRG-S0851 (see Figure~\ref{fig:srcrec-GALFIT}).

\subsection{Results of the Unresolved Photometric Analysis\label{subsec:prospector}}

\begin{figure}[!t]
\centering
\includegraphics[width=1.0\columnwidth]{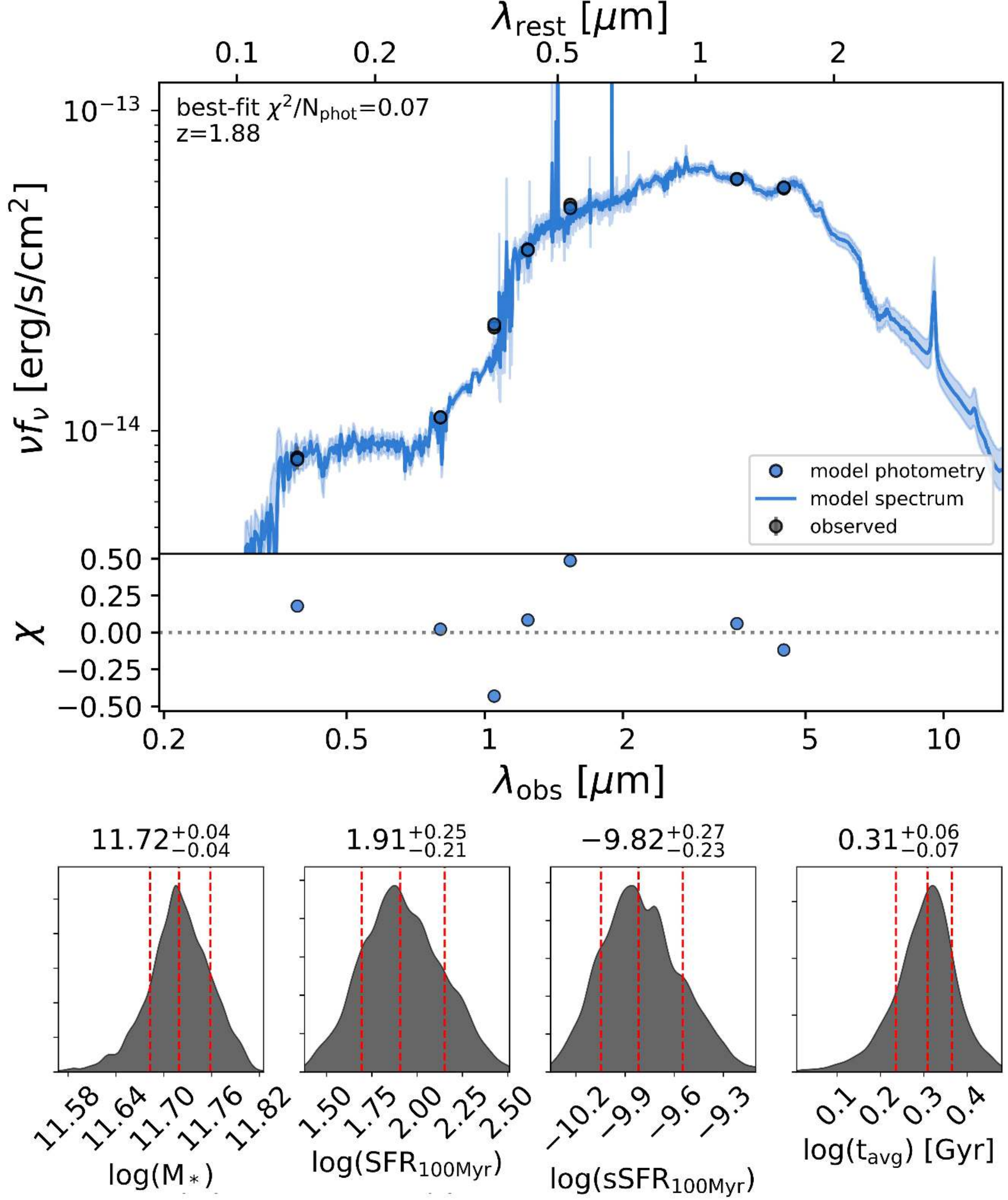}
\caption{The unresolved global stellar populations fit, which is obtained using \texttt{Prospector-$\alpha$}, to 5 \emph{HST} bands and 2 \emph{Spitzer} IRAC channels. Stellar mass and star-formation rate (SFR) are not corrected for strong gravitational lensing effect in this plot. \label{fig:0851-E-prospector}}
\end{figure}

\begin{table}
\begin{center}
\begin{tabular}{|c|c|c|c|c|}
 \hline
{} & \texttt{dust1} & \texttt{dust2} & \texttt{dust\_index} & $\log Z/Z_\odot$ \\
\hline\hline
Global & $0.69^{+0.72}_{-0.43}$ & $0.73^{+0.43}_{-0.39}$ & $0.15^{+0.22}_{-0.38}$ & $-0.04^{+0.21}_{-0.44}$\\ \hline
$<$-1.4 kpc & $0.66^{+0.71}_{-0.41}$ & $0.87^{+1.06}_{-0.65}$ & $-0.57^{+0.46}_{-0.38}$ & $-0.24^{+0.40}_{-0.84}$ \\ \hline 
-1.0 kpc & $0.83^{+1.34}_{-0.64}$ & $0.87^{+1.06}_{-0.65}$ & $0.05^{+0.29}_{-0.54}$ & $-0.17^{+0.33}_{-0.52}$ \\ \hline
-0.5 kpc & $0.9^{+1.44}_{-0.78}$ & $0.84^{+1.07}_{-0.68}$ & $0.15^{+0.22}_{-0.75}$ & $-0.14^{+0.3}_{-0.49}$ \\ \hline
Center & $1.25^{+1.46}_{-1.18}$ & $1.22^{+1.05}_{-1.15}$ & $0.32^{+0.08}_{-1.05}$ & $-0.11^{+0.28}_{-0.42}$ \\ \hline
+0.8 kpc & $1.02^{+1.53}_{-0.95}$ & $1.02^{+1.14}_{-0.93}$ & $0.24^{+0.15}_{-0.58}$ & $-0.17^{+0.33}_{-0.46}$\\ \hline
+1.2 kpc  & $0.93^{+1.28}_{-0.73}$ & $0.96^{+0.98}_{-0.72}$ & $-0.04^{+0.39}_{-0.76}$ & $-0.15^{+0.32}_{-0.59}$ \\ \hline
$>$+1.5 kpc & $0.71^{+1.3}_{-0.6}$ & $0.69^{+1.03}_{-0.57}$ & $-0.28^{+0.62}_{-0.66}$ & $-0.08^{+0.24}_{-0.52}$\\
\hline
\end{tabular}
\caption{The best-fit values of dust and metallicity for the global photometric measurements of MRG-S0851 (5 \emph{HST} + 2 \emph{Spitzer} bands) as well as spatially resolved bins (5 \emph{HST} bands only), as defined in Section \ref{subsec:def-bins} and shown in Figure \ref{fig:bins}, obtained by \texttt{Prospector-$\alpha$}. \citet{kriek2013} dust model is assumed here. In this model, \texttt{dust1} and \texttt{dust2} are the parameters controlling the strength of dust attenuation for the stellar populations younger and older than $10^7$ years, respectively, and  \texttt{dust\_index} controls the slope of dust attenuation curve. \label{tab:dust-metal}}
\end{center}
\end{table}

The unresolved global stellar population fit to 7 photometric bands for MRG-S0851, obtained using \texttt{Prospector-$\alpha$} as discussed in Section \ref{subsec:dust-metal-prior}, is shown in Figure \ref{fig:0851-E-prospector}. The \texttt{Prospector-$\alpha$} fit yields a mass-weighted average age of $2.0\pm 0.3$ Gyr for the unresolved stellar populations, and a specific star-formation rate (sSFR) of $\log \mathrm{sSFR}_{\mathrm{100 Myr}}/[\mathrm{yr^{-1}}] = -9.8^{+0.3}_{-0.2}$, independent of the lensing magnification.

The demagnified stellar mass of MRG-S0851 is estimated to be $\log M_* / M_\odot = 10.96\pm 0.04$ and the demagnified SFR is constrained to be $\log \mathrm{SFR}_{100\mathrm{Myr}}/[M_\odot \mathrm{yr^{-1}}]=1.2^{+0.3}_{-0.2}$, correcting for strong gravitational lensing magnification of $\mu _{\mathrm{E3}}=5.7^{+0.4}_{-0.2}$. The propagated uncertainty of gravitational magnification on these parameters is negligible relative to the uncertainty of the photometric fit. As discussed in Section \ref{subsec:prospector-phot-fit}, these results are obtained using photometric data alone, and we next jointly fit spectro-photometric data to obtain constraints on the age of the stellar populations.

The dust and metallicity values for MRG-S0851 are reported in Table \ref{tab:dust-metal}. The best fit \texttt{dust2} parameter of MRG-S0851 implies an $A_V$ of $\sim0.8$; however, as \texttt{dust\_index}$\sim$0.15, the attenuation curve will be flatter than the Calzetti law, and we have less UV attenuation and more near-IR attenuation comparably, as discussed in Section \ref{subsec:prospector-phot-fit}. The measured metallicity value has a large uncertainty, consistent with both a sub-solar and a solar metallicity within the error bars. A tighter constraint is necessary to speculate about the future chemical enrichment of MRG-S0851.

\subsection{Results of the Spectro-Photometric Analysis for the case study of MRG-S0851 \label{sec:age-grad}} 

In this Section, we apply the methodology described in Section \ref{subsec:method} to MRG-S0851. We include the photometric data obtained from five WFC3 filter images ($\mathrm{H}_{\mathrm{F160W}}$, $\mathrm{J}_{\mathrm{F125W}}$, $\mathrm{Y}_{\mathrm{F105W}}$, $\mathrm{I}_{\mathrm{F814W}}$ and $\mathrm{U}_{\mathrm{F390W}}$) along with the spectroscopic data of WFC3-G141 in our joint-fitting. For the unresolved analysis, we explicitly include the two \emph{Spitzer} IRAC bands and for the resolved analysis, we require that the sum of spatial bins' IRAC fluxes matches two unresolved IRAC fluxes as an additional constraint.

\begin{figure}
\centering
\includegraphics[width=1.0\columnwidth]{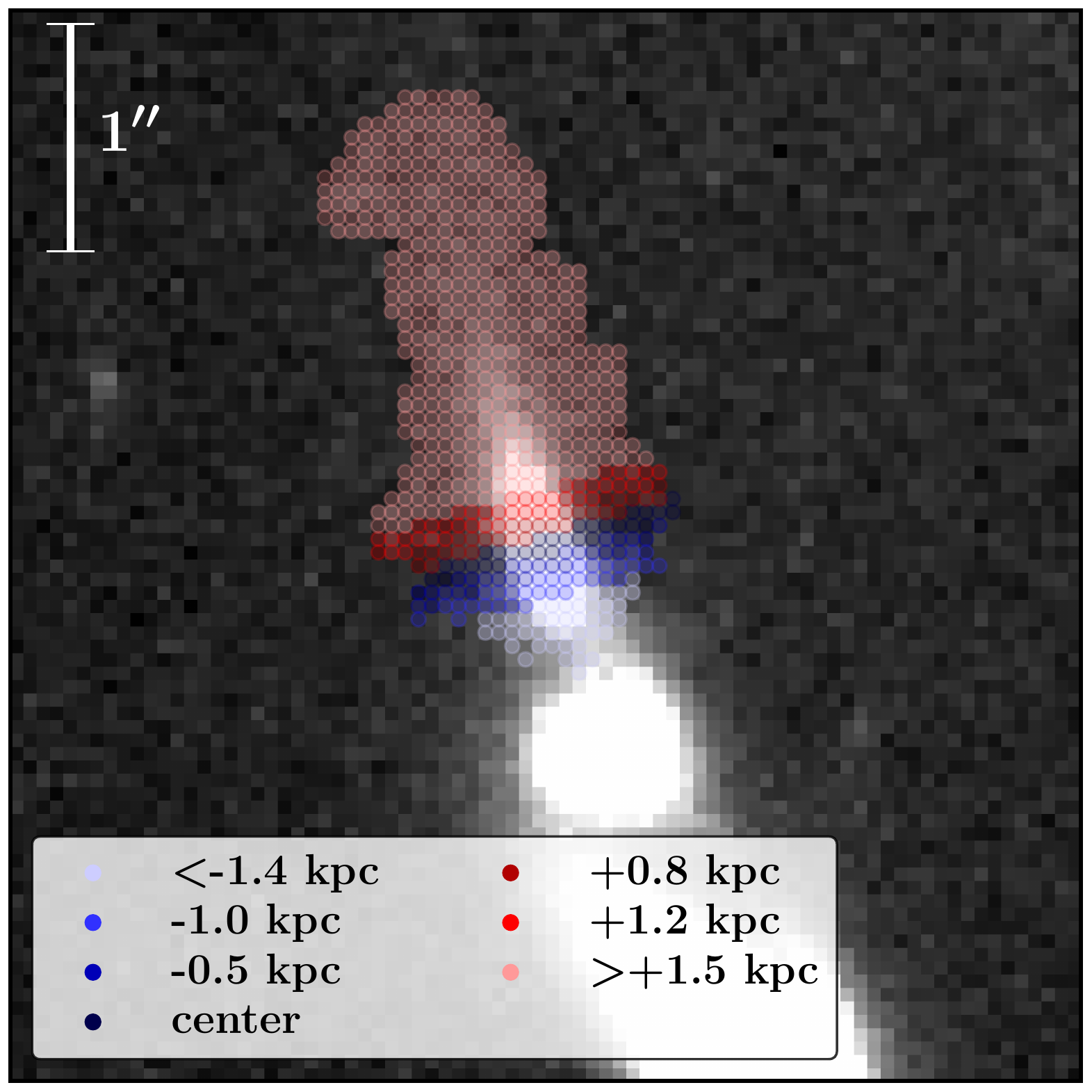}
\caption{$\mathrm{H_{F160W}}$ image of MRG-S0851, showing the pixels used to define the spatial bins (see Section \ref{subsec:def-bins}).  While this figure shows the image-plane location, their relative location within the source-plane of MRG-S0851 is noted in the legend.  \label{fig:bins}}
\end{figure}

Figure \ref{fig:bins} shows the 7 spatial bins for each grism exposure obtained following the recipe of Section \ref{subsec:def-bins}. We explicitly check that the center bin also includes the peak of the $\mathrm{U}_{\mathrm{F390W}}$ light profile.

Each pixel in the image is $0\farcs06$, so the central bins range from $0\farcs18$ to $0\farcs24$ wide. Using our lensing model (see Appendix \ref{sec:cluster-lens}), the bins are centered at distances of  $<$-1.4 kpc, -1.0 kpc, -0.5 kpc, 0, +0.8 kpc, +1.2 kpc, and $>$+1.5 kpc. The 5 central bins therefore span $2.9 \, \mathrm{kpc}$ in total in the source plane, which is comparable to the half-light diameter of $\simeq$3.4 kpc measured in Appendix \ref{subsec:size}. We therefore probe the age gradient of the stellar populations in the inner radius of $\simeq 1.7 \,  \mathrm{kpc}$ of MRG-S0851 at an average spatial resolution of $\simeq 0.6 \, \mathrm{kpc}$.

\begin{figure*}[!]
\gridline{\fig{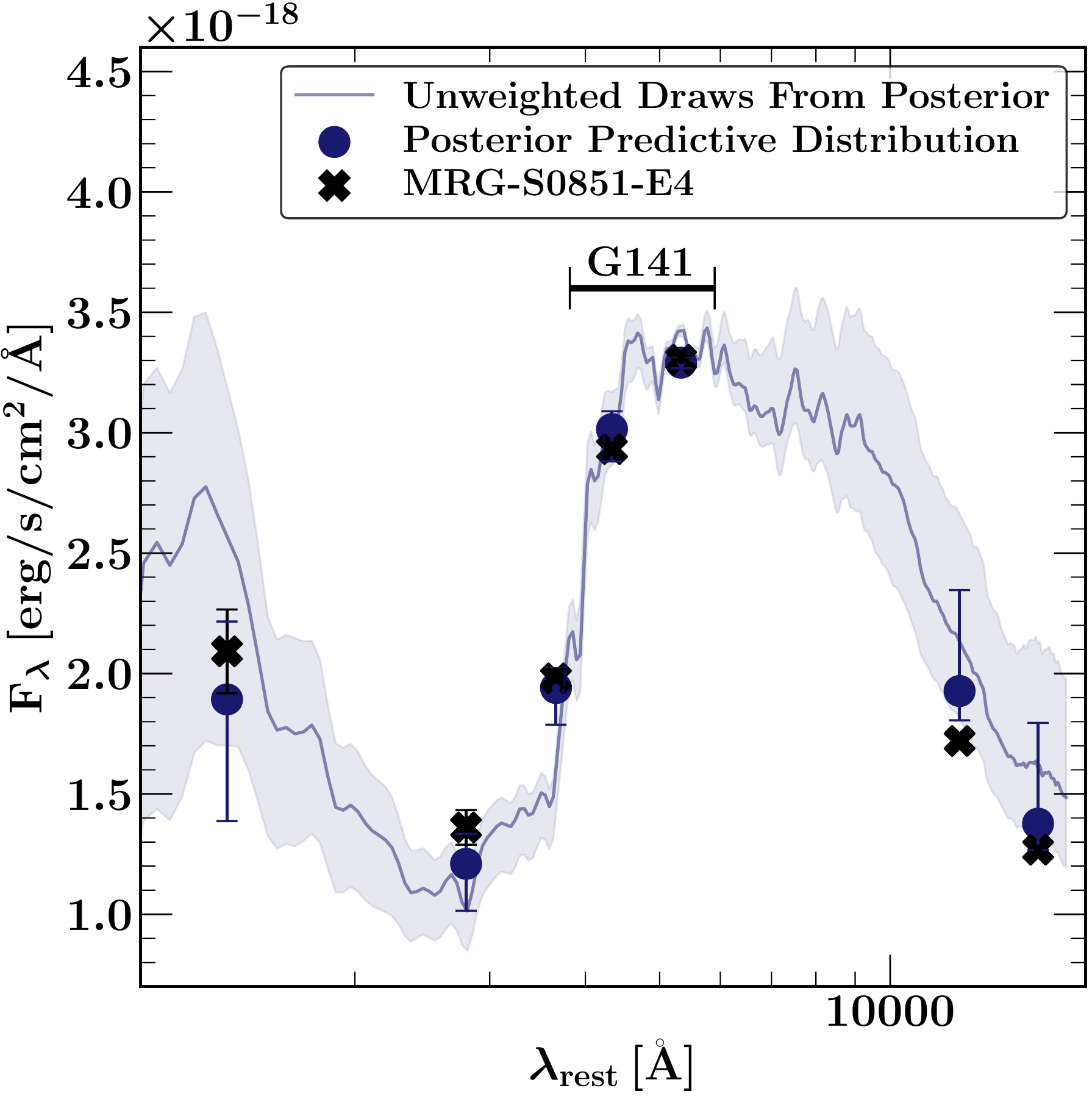}{0.415\textwidth}{}
\fig{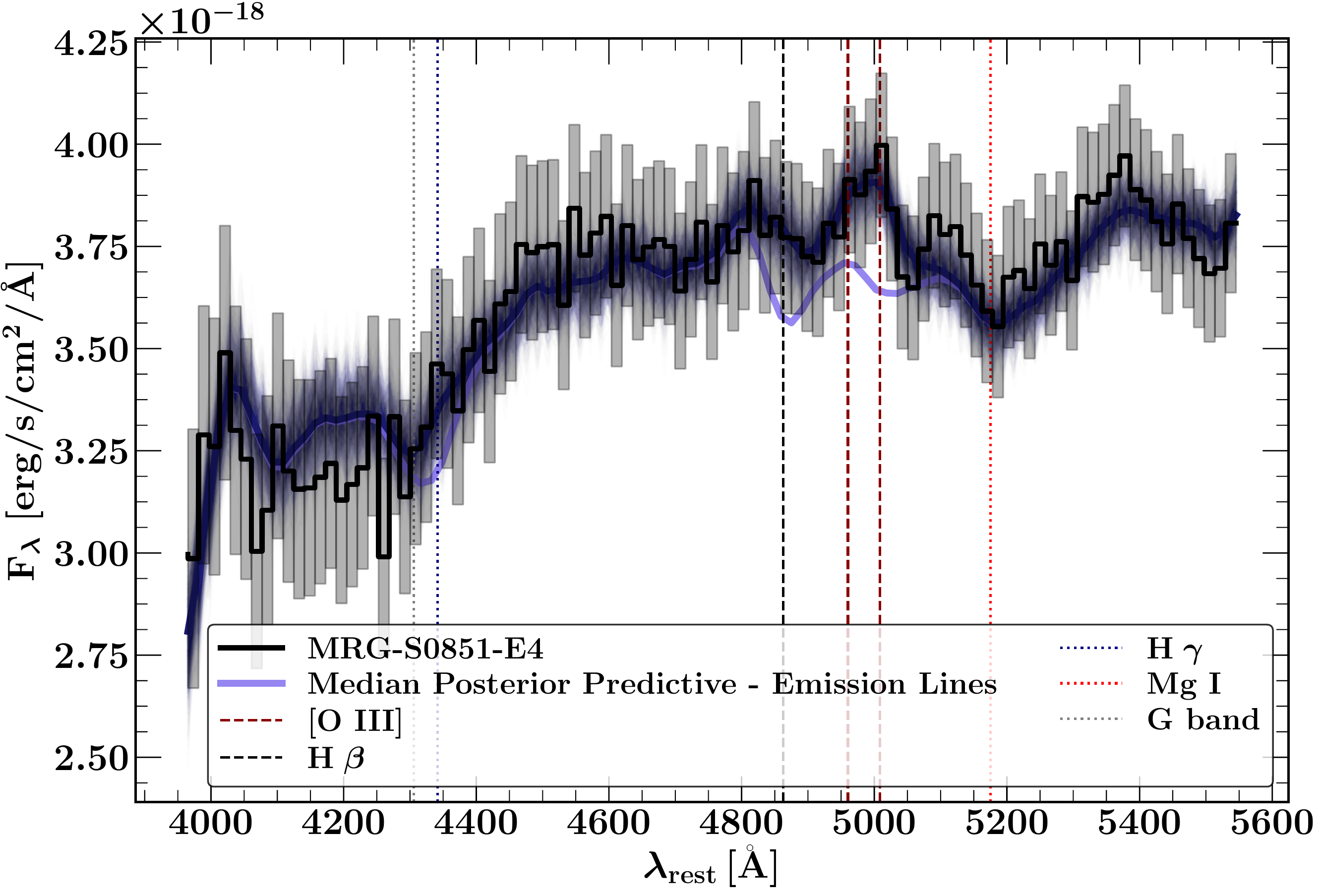}{0.617\textwidth}{}}
\vspace{-0.25in}
\caption{\textbf{On the left:} The global spectral energy distribution (SED) of MRG-S0851. The black crosses are the \emph{HST} and \emph{Spitzer} photometric data; the blue circles are the posterior predictive distribution, and dark blue curve and its light blue shade are the unweighted median and 16th-84th percentiles of the draws from posterior (i.e., unlike the blue circles, they are not weighted by the weights of the Dirichlet process). \textbf{On the right:} The global extracted 1D grism spectra of MRG-S0851, following \citep{horne1986} for the 1D extraction. Various spectral features in the G141 bandpass at the redshift of MRG-S0851, are shown by vertical lines. The data is shown in black, with the grey vertical bars as 1$\sigma$ errors , and the dark blue curves are draws from the posterior predictive distribution. We smooth both the data and models using the \texttt{gaussian\_filter} method of \texttt{scipy} with $\sigma = 0.3$. We also plot the full median model with a darker blue line and the median continuum-only model removing emission lines with a lighter blue line. \label{fig:unresolved-SDSS-J0851-E}}
\end{figure*}

The results for the unresolved analysis of the stellar populations of MRG-S0851 are shown in Figure \ref{fig:unresolved-SDSS-J0851-E}, with the 1D grism spectra in the right panel, extracted following \citet{horne1986}\footnote{The 1D extraction is only performed to make the final plot at the end of the analysis; we fit and analyze the grism spectra in its native 2D space.}. The spectral energy distribution (SED) of the unresolved stellar populations can be found in the left panel of Figure \ref{fig:unresolved-SDSS-J0851-E}. MRG-S0851 has significant flux in the rest-frame UV, which is sampled by the photometric filter $\mathrm{U}_{\mathrm{F390W}}$. This rest-frame UV flux indicates a recent star-formation activity detected in the last $\sim$100 Myr of evolution in MRG-S0851, which is overall $\sim$1.4 dex below star-forming sequence of \citet{whitaker2014} and $\sim$1.1 dex below star-forming sequence of \citet{leja2018} (based on the \texttt{requiem2d} joint fit).  We will discuss this result in depth in a subsequent paper (Akhshik et al. in prep).  

\begin{figure*}[!th]
\gridline{\fig{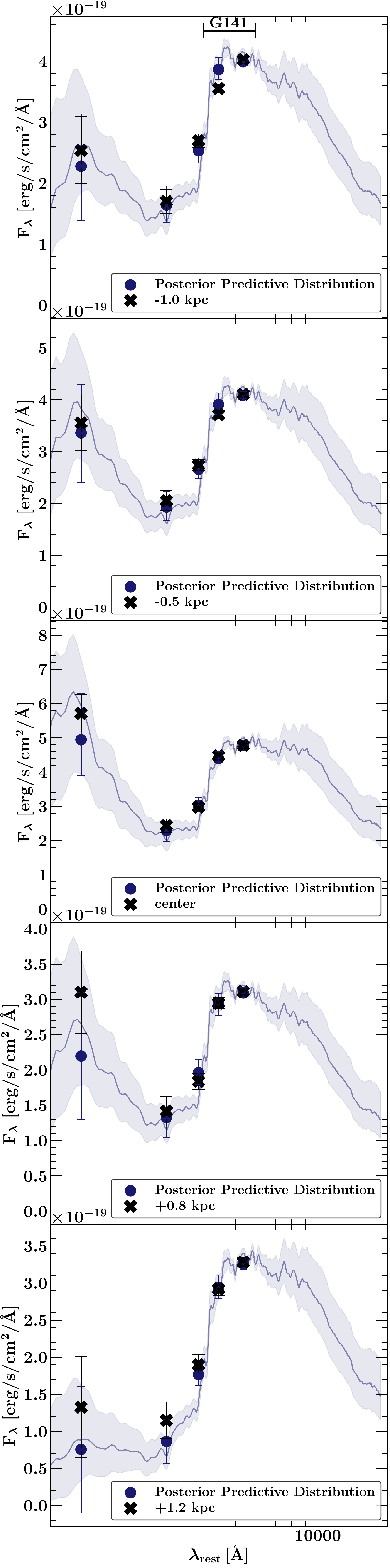}{0.28\textwidth}{}
\hspace{-1.5in}\fig{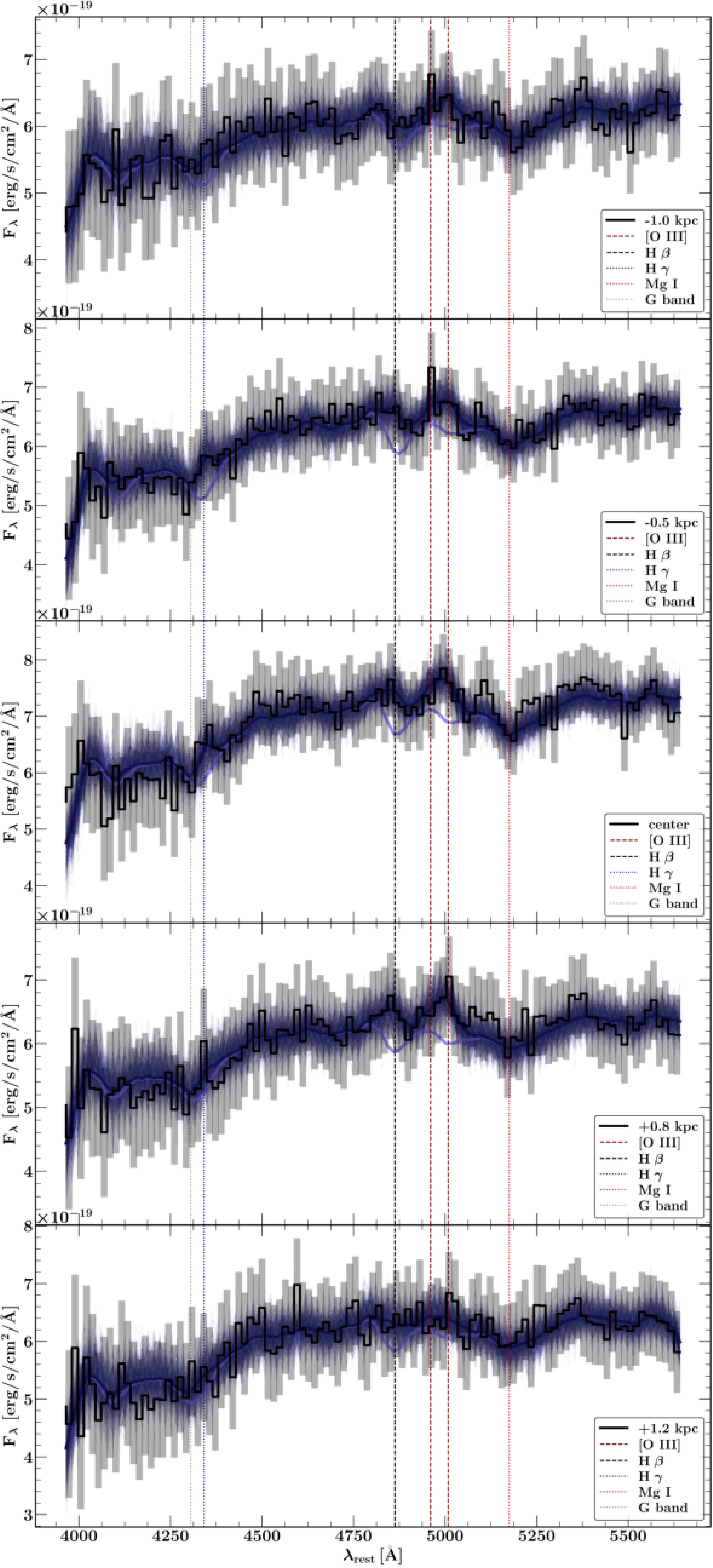}{0.51\textwidth}{}}
\vspace{-0.3in}
\caption{The resolved extracted 1D grism spectra and the SED of MRG-S0851, obtained using the global SSP prior for dust and metallicity (Section \ref{subsec:dust-metal-prior}). We smooth both the data and the posterior predictive distribution using the \texttt{gaussian\_filter} method of \texttt{scipy} with $\sigma = 0.3$. The symbols and the 1D extraction method are the same as the Figure \ref{fig:unresolved-SDSS-J0851-E}.  \label{fig:resolved-SDSS-J0851-E}}
\end{figure*}

\begin{figure}
\centering
\includegraphics[width=0.95\columnwidth]{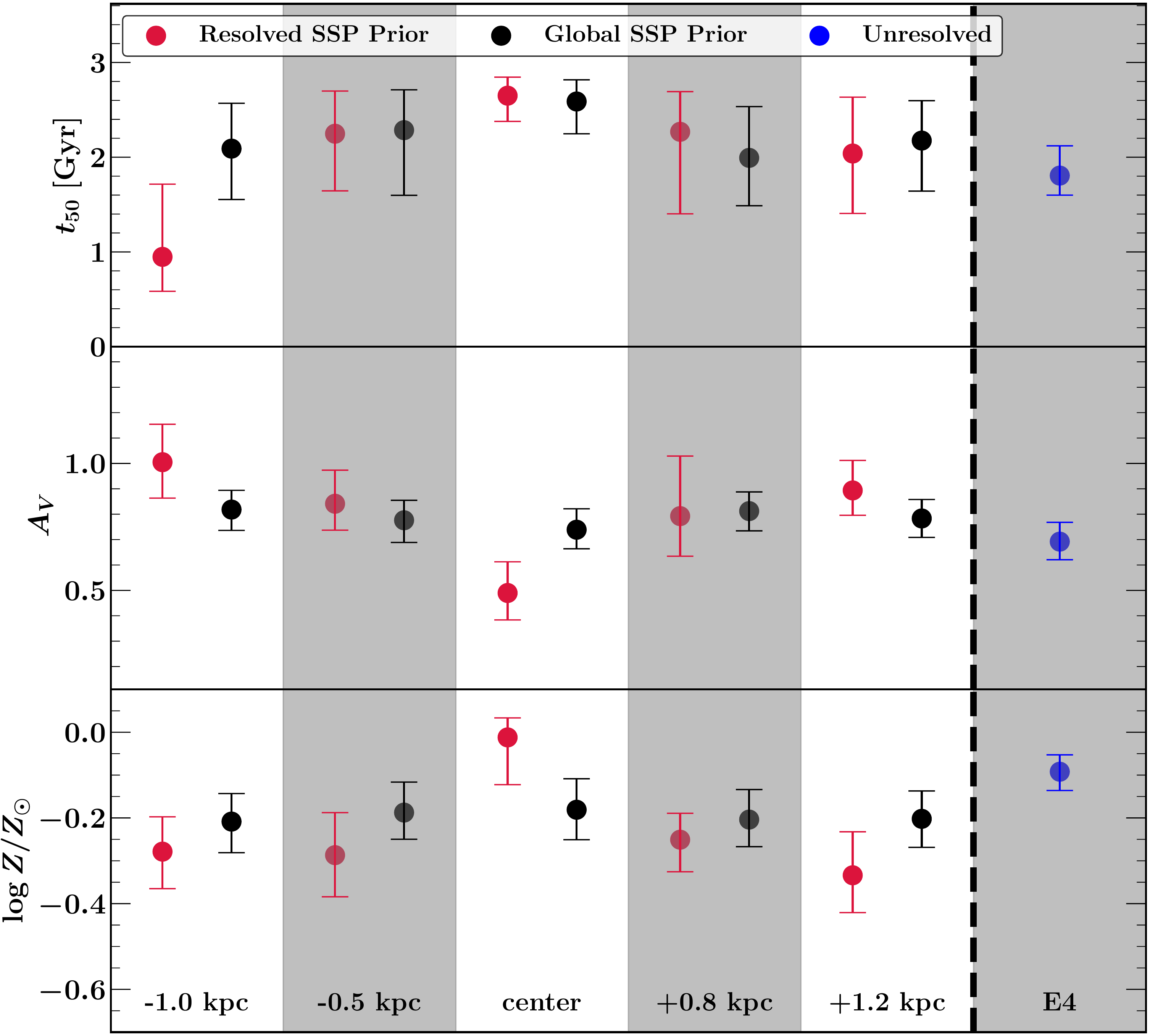}
\caption{The resolved median ages, $A_V$, and metallicities of central spatial bins (left) as well as the unresolved global values MRG-S0851 (right). For the spatial bins, we plot two sets of mass-weighted average ages obtained using two different dust/metallicity priors (Section \ref{subsec:dust-metal-prior}). One set is obtained by using the resolved \texttt{Prospector-$\alpha$} dust/metallicity posteriors as priors to generate SSPs in the joint fitting (dubbed Resolved SSP Prior), and the other is obtained by using the global \texttt{Prospector-$\alpha$} dust/metallicity posterior as a prior, generating SSPs for all spatially resolved bins (dubbed Global SSP Prior). The blue circles in the right panel is the global values of MRG-S0851, as measured in the unresolved analysis. \label{fig:age-grad}}
\end{figure}

We measure a global median mass-weighted age of $1.8_{-0.2}^{+0.3}$ Gyr with a  corresponding global light-weighted age of $1.4^{+0.2}_{-0.1}$ Gyr from the joint-fitting. The results for the 5 central bins of MRG-S0851 are also shown in Figure \ref{fig:resolved-SDSS-J0851-E}. Figure \ref{fig:age-grad} shows the median mass-weighted age for both the resolved and unresolved stellar populations. 
Through our spatially-resolved analysis, we were not able to robustly constrain the ages of the two outer spatially-resolved bins at $<$-1.4 kpc and $>$1.5 kpc, as our two choices of dust/metallicity SSP priors (see the two last columns of Table \ref{tab:fits}) lead to different ages. This is likely the result of the significant contamination in these outer grism spectra by the cluster member galaxies and other nearby objects, yielding an average $\mathrm{S}_{\mathrm{grism}}/\mathrm{N}_{\mathrm{contam}}$ ratio of $\sim$0.2 and 1.1 per pixel for these bins, where $\mathrm{S}_{\mathrm{grism}}$ is the flux of MRG-S0851 and $\mathrm{N}_{\mathrm{contam}}$ is the flux of other nearby objects in each pixel.  For reference, the central bin has $\mathrm{S}_{\mathrm{grism}}/\mathrm{N}_{\mathrm{contam}}\sim 22.8$ per pixel. Our analysis suggests that MRG-S0851 has a flat age gradient within the inner 3 kpc core, with a circularized effective radius of $1.7^{+0.3}_{-0.1}$kpc in the source plane (see Appendix \ref{sec:whole arc}, for size measurements), it is similarly compact relative to the population of quiescent galaxies with the same stellar mass and redshift \citep[e.g.,][]{vanderwel2014}. With a median age of 1.8 Gyr, it seems to be older than similar quiescent galaxies selected from the Illustris simulation, that have $t_{50}=1.1^{+0.3}_{-0.2}$ Gyr and $\log M_*/M_\odot=11.03^{+0.30}_{-0.37}$ (see Figure \ref{fig:mock_test2}). MRG-S0851 is therefore consistent with an early formation scenario \citep[e.g.,][]{williams2014,wellons2015}. The recovered SFH gradients of MRG-S0851 will be discussed in further depth in a follow-up paper (Akhshik et al. in prep).

We estimate the average mass-weighted age of $1.9^{+0.2}_{-0.1}$~Gyr from our joint fit with \texttt{requiem2d}, consistent within the 1$\sigma$ uncertainty interval with the estimate of the average mass-weighted age from \texttt{Prospector-$
\alpha$} ($2.0\pm0.3$~Gyr; Figure~\ref{fig:0851-E-prospector}) and the median mass-weighted age obtained from the joint fit ($1.8^{+0.3}_{-0.2}$~Gyr). We estimate a stellar mass of $\log M_*/M_\odot=11.02\pm0.04$ from the joint fit, consistent with the \texttt{Prospector-$\alpha$} estimate within 1$\sigma$ (Section \ref{subsec:prospector}). Finally, the sSFR is estimated to be $\log \mathrm{sSFR}_{\mathrm{100 Myr}}/[\mathrm{yr^{-1}}]=-10.32_{-0.05}^{+0.07}$ from the spectro-photometric fit. It is $\sim$0.5 dex lower than the sSFR estimate of \texttt{Prospector-$\alpha$} from photometry only. We discuss this discrepancy in Section \ref{sec:summary}.

In the G141 bandpass at the redshift of MRG-S0851 z=1.88, we observe a few spectral features that are sensitive to the age of stellar populations. Most notably, we sample the $\mathrm{4000 \AA}$ break, $\mathrm{Mg \, I}$ and the Balmer lines $\mathrm{H\beta}$, $\mathrm{H\gamma}$, and $\mathrm{H\delta}$. As the Balmer lines appear to be filled by emission, they are not expected to drive the age fit significantly. We detect $[\mathrm{O\, III}]$ emission lines at the rest-frame wavelengths of $5008.2 \mathrm{\AA}$ and $4960.3 \mathrm{\AA}$ (to be discussed further in Section \ref{subsec:emission-lines}). 

As discussed in Section \ref{subsec:model-elements}, we add the emission line templates using \texttt{Grizli} together with the SSPs that do not include nebular emission lines that are generated by FSPS. Another option is to instead use FSPS to generate nebular emission lines. To test this alternative approach, we use the \texttt{Prospector-$\alpha$} posterior of gas-phase metallicity along with the stellar metallicity and dust parameters to add nebular emission lines, specifically adding more templates between 1-10 Myr in logarithmic lookback time steps. We confirm that the resulting global and resolved ages are all consistent in both approaches within 1$\sigma$ statistical uncertainty, except for bins at $\leq$-1.4kpc and $\geq$1.5kpc due to contamination. The global mass-weighted median age using FSPS templates that include nebular emission lines is constrained to be $1.9\pm0.2$ Gyr.

When accounting for the effects of dust and metallicity on the SSPs of resolved stellar populations of MRG-S0851, the distribution of resolved ages may change for different priors (Figure \ref{fig:age-grad}, comparing black and red points; Also see Section \ref{subsec:joint-fit-prior} for further discussion of these priors). We can not calculate the resolved photometry for the two \emph{Spitzer} IRAC channels, and as a result, neither the resolved joint-fit nor the resolved \texttt{Prospector}-$\alpha$ fit fully includes resolved rest-frame near-IR photometry\footnote{As a reminder, in our spatially-resolved joint-fit, we require that the sum of the predicted resolved IRAC photometric fluxes in our model match the observed global values (Section \ref{subsec:data-prep})}. Therefore, the priors on SSPs, that control dust and metallicity, may play a significant role and affect the weights $\mathbf{x}$ of SSPs. While this effect is not large for 5 central bins, it seems to be more significant for the outer bins at $\leq$-1.4 kpc and $\geq$1.5 kpc, where the SSP priors strongly driving the fit and yielding different results for the ages of these bins as the photometry error bars are larger and grism spectroscopy is significantly contaminated.

\subsection{Emission-line Diagnostics of MRG-S0851 \label{subsec:emission-lines}}

We fit for the fluxes of emission lines, as described in Sections \ref{subsec:model-elements} and \ref{sec:age-grad}. The statistical significance of the spatially-unresolved emission lines is estimated to be $\sim 9 \sigma$ ($[\mathrm{O III}]$), $\sim 5 \sigma$ ($\mathrm{H}\beta$), $\sim 3 \sigma$ ($\mathrm{H}\gamma$), and $\sim 1 \sigma$ ($\mathrm{H}\delta$) from the corresponding Markov Chains, 
and we therefore detect $[\mathrm{O III}]$, $\mathrm{H}\beta$ and $\mathrm{H}\gamma$ with $\gtrapprox 3\sigma$ statistical certainty. 

$[\mathrm{O III}]$ is detected in three centrals bins ($>3\sigma$). We constrain the global $\log \left( [\mathrm{O III}]/\mathrm{H}\beta \right)$ to be $0.24^{+0.10}_{-0.08}$. Taken together, it is likely that the emission lines originate from star-formation activity.  It may however be that MRG-S0851 has a high $[\mathrm{N II}]/\mathrm{H}\alpha$ ratio, which combined with low $[\mathrm{O III}]/\mathrm{H}\beta$, would favor an AGN \citep{BPT1981} or a low-ionization emission-line region \citep[e.g. ][]{heckman1980, belfiore2016}. Unfortunately, both $[\mathrm{N II}]$ and $\mathrm{H}\alpha$ are outside of the WFC3/G141 bandpass (and unresolveable) for MRG-S0851 to examine this scenario. 

In the spatially resolved analysis, we also fit for the flux of $[\mathrm{O III}]$, $\mathrm{H}\beta$, $\mathrm{H}\gamma$ and even $\mathrm{H}\delta$ emission lines in the WFC3/G141 bandpass, but the weakness of these emission lines combined with the low spectral resolution of WFC3/G141 do not allow us to constrain any potential gradients in $[\mathrm{O III}]/\mathrm{H}\beta$, or statistically significant Balmer decrements.

\section{SUMMARY AND DISCUSSION \label{sec:summary}}

We present a methodology to jointly fit \emph{HST} grism spectroscopy with \emph{HST} and \emph{Spitzer} photometry with the new \texttt{requiem2d} package to constrain the age and SFH of stellar populations of galaxies. Our fitting method includes two steps, a preliminary fit with \texttt{Prospector}-$\alpha$ to photometric data, and a subsequent joint spectro-photometric fit using a linear combination of SSPs, generated by drawing from the \texttt{Prospector}-$\alpha$ posterior to constrain ages and SFHs. Our presentation here is tuned to \emph{HST} grism spectroscopy and \emph{HST} and \emph{Spitzer} photometry, however, the core statistical model (Table \ref{tab:regresssion} and Figure \ref{fig:model-plate}) can be applied to all well-calibrated spectroscopic and photometric data. The beta version release of the \texttt{requiem2d} code accompanies this manuscript, accessible through a public GitHub repository\footnote{https://github.com/makhshik/requiem2d}.

We test our methodology using a sample of massive $\log M_*/M_\odot \geq 10.6$ quiescent galaxies at $z=2$ selected from the Illustris simulation, showing that the median mass-weighted age as well as the general trends of SFH can be recovered with no noticeable biases. We also use this method in a pilot study to analyze the spatially-resolved stellar populations of a lensed massive red galaxy, MRG-S0851, using photometric data taken using \emph{HST}/WFC3 broadband filters which cover the rest-frame UV to optical, \emph{HST}/WFC3 G141 grism data, and \emph{Spitzer} IRAC channels 1 and 2 data. With grism spectroscopy, we constrain the redshift to be $z=1.883 \pm 0.001$ by fitting MRG-S0851 using \texttt{Grizli}. By constructing a consistent lensing model, we correct the stellar mass for the effects of strong gravitational lensing, constraining the stellar mass to be $\log M_* / M_\odot = 11.02\pm 0.04$ from our joint-fit with \texttt{requiem2d}. The circularized effective radius is measured to be $r_c=1.7^{+0.3}_{-0.1}$ kpc ($0\farcs21 \pm 0\farcs02$) in the source plane (see Appendix \ref{subsec:size}). We fit the global dust and metallicity using \texttt{Prospector}-$\alpha$, with the results reported in Table \ref{tab:dust-metal}. From a joint spectro-photometric analysis, we find that the unresolved stellar populations have a global median mass-weighted age of $1.8_{-0.2}^{+0.3}$ Gyr, a global light-weighted age of $1.4^{+0.2}_{-0.1}$ Gyr, and a global specific star formation rate of $\log \mathrm{sSFR}_{\mathrm{100 Myr}}/[\mathrm{yr^{-1}}]=-10.32_{-0.05}^{+0.07}$. The sSFR obtained using \texttt{requiem2d} from the joint-fit is 0.5 dex lower than the \texttt{Prospector-$\alpha$} sSFR from a fit to photometry alone (Figure~\ref{fig:0851-E-prospector}). We confirm that the difference is a result of adding grism spectroscopy, and fitting photometry alone using \texttt{requiem2d} yields consistent results with \texttt{Prospector-$\alpha$}. In fact, by adding grism spectroscopy, that lacks the relatively strong Balmer absorption lines of ages $\sim$0.1-1 Gyr, the SFR in this range of lookback time decreases while the SFR  at $\geq$1-2 Gyr increases, leading to an overall lower sSFR in the last $\sim$100 Myr of evolution.  The change in SFH in the joint spectro-photometric versus photometry-only fit therefore mostly affects lookback-time ranges of 0.1-2 Gyr. However, this change in the overall shape of the inferred SFH does not change the median/average mass-weighted ages of MRG-S0851 significantly, as these ages are dominated by the older SSP templates with lookback-times of $\gtrapprox2$~Gyr. 

Leveraging the strong gravitational lensing magnification, we define 7 spatial bins to study the spatially resolved stellar populations and to measure the age gradient (Figure \ref{fig:age-grad}). We defer the discussion of the SFH gradient to a subsequent paper. At face value, the flat age gradient combined with a relatively old global age and compact size favors an early formation scenario for MRG-S0851 \citep[e.g.,][]{williams2014,wellons2015}

Emission-line diagnostics in the \emph{HST}/WFC3 G141 bandpass at the redshift of MRG-S0851 disfavor the presence of an AGN at the center of the galaxy, as $[\mathrm{O III}]$ emission line is not centrally concentrated and the ratio of $[\mathrm{OIII}]/\mathrm{H}\beta$ is not high. While the $[\mathrm{OIII}]/\mathrm{H}\beta$ ratio is consistent with the star formation activity, we caution that we need to observe $\mathrm{H}\alpha$ and $[\mathrm{NII}]$ to rule out AGN or low-ionization emission-line regions. These emission lines can be studied easily in the near future with the NIRSpec/Integrated Field Unit (IFU) of the \emph{James Webb Space Telescope (JWST)}. These lines are also accessible to ground-based spectroscopy using Keck/MOSFIRE, albeit at a lower resolution than \emph{JWST}. Observations of $\mathrm{H}\alpha$ will also be helpful to constrain the instantaneous star-formation rate.

MRG-S0851 is the first target in the ongoing REQUIEM galaxy survey, which includes grism spectroscopy using \emph{HST}/WFC3 G141 as a part of the HST-GO-15663 ongoing program, for a sample of 8 strongly lensed quiescent galaxies spanning a redshift range of $1.6<z<2.9$ and a stellar mass range of $10.4<\log M_*/M_\odot<11.7$ (HST-GO-15633). The analysis of the spatially resolved stellar populations of the rest of the REQUIEM targets will follow the framework developed herein.

\section*{acknowledgements}
We would like to thank the anonymous referee for sharing valuable comments and questions that helped us improve the draft significantly. We gratefully acknowledge support by NASA under award No 80NSSC19K1418, HST-GO-14622, and HST-GO-15663. This work is based on observations made with the NASA/ESA Hubble Space Telescope, obtained at the Space Telescope Science Institute, which is operated by the Association of Universities for Research in Astronomy, Inc., under NASA contract NAS 5-26555. Financial support for M.A. and K.W. is gratefully acknowledged. K.W. wishes to acknowledge funding from the Alfred P. Sloan Foundation.
S.T. acknowledges support from the ERC Consolidator Grant funding scheme (project ConTExt, grant No. 648179). The Cosmic Dawn Center is funded by the Danish National Research Foundation. C.C.W acknowledges support from the National Science Foundation Astronomy and Astrophysics Fellowship grant AST-1701546. The Dunlap Institute is funded through an endowment established by the David Dunlap family and the University of Toronto.

\bibliography{master}

\appendix

\section{GRAVITATIONAL LENS MODELING OF MRG-S0851 \label{sec:cluster-lens}}

We highlight the key aspects of the MRG-S0851 lens model. This lens model is used to transform information to the source plane and interpret the results of stellar population analyses. 

\subsection{Methodology of Lensing Mass Models}
\label{sect:lensmethod}

\begin{figure*}
\centering
\gridline{\fig{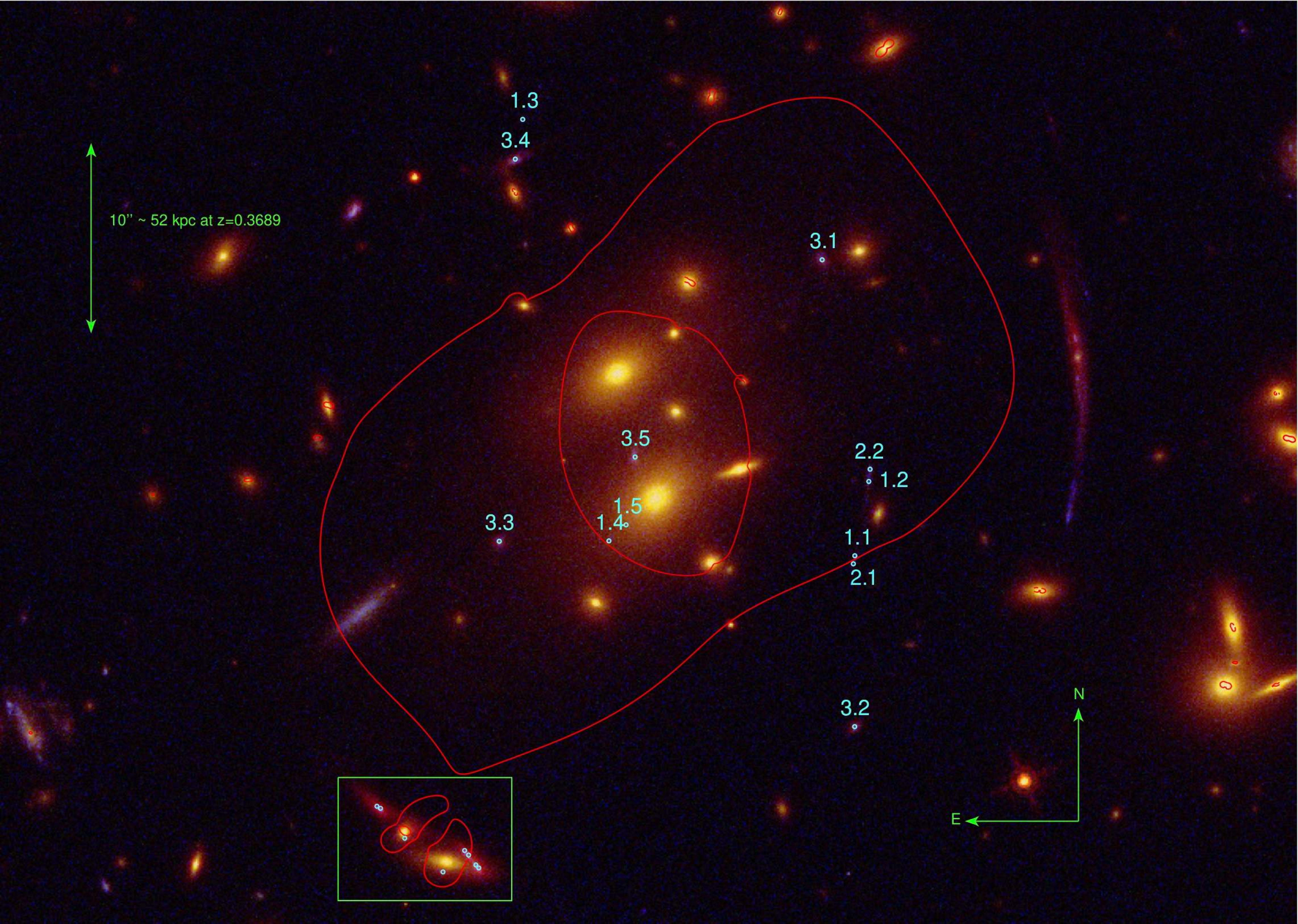}{1.0\textwidth}{}}
\gridline{\fig{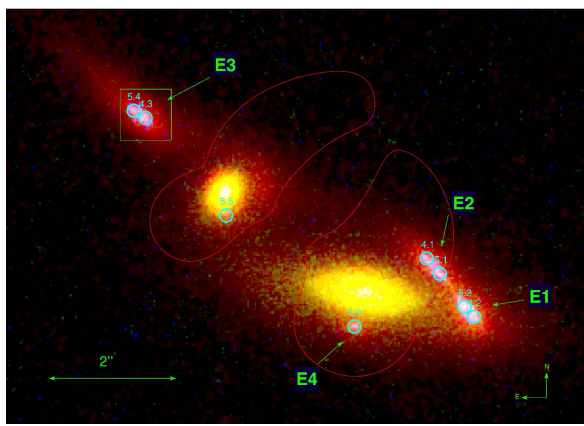}{0.43\textwidth}{}
\fig{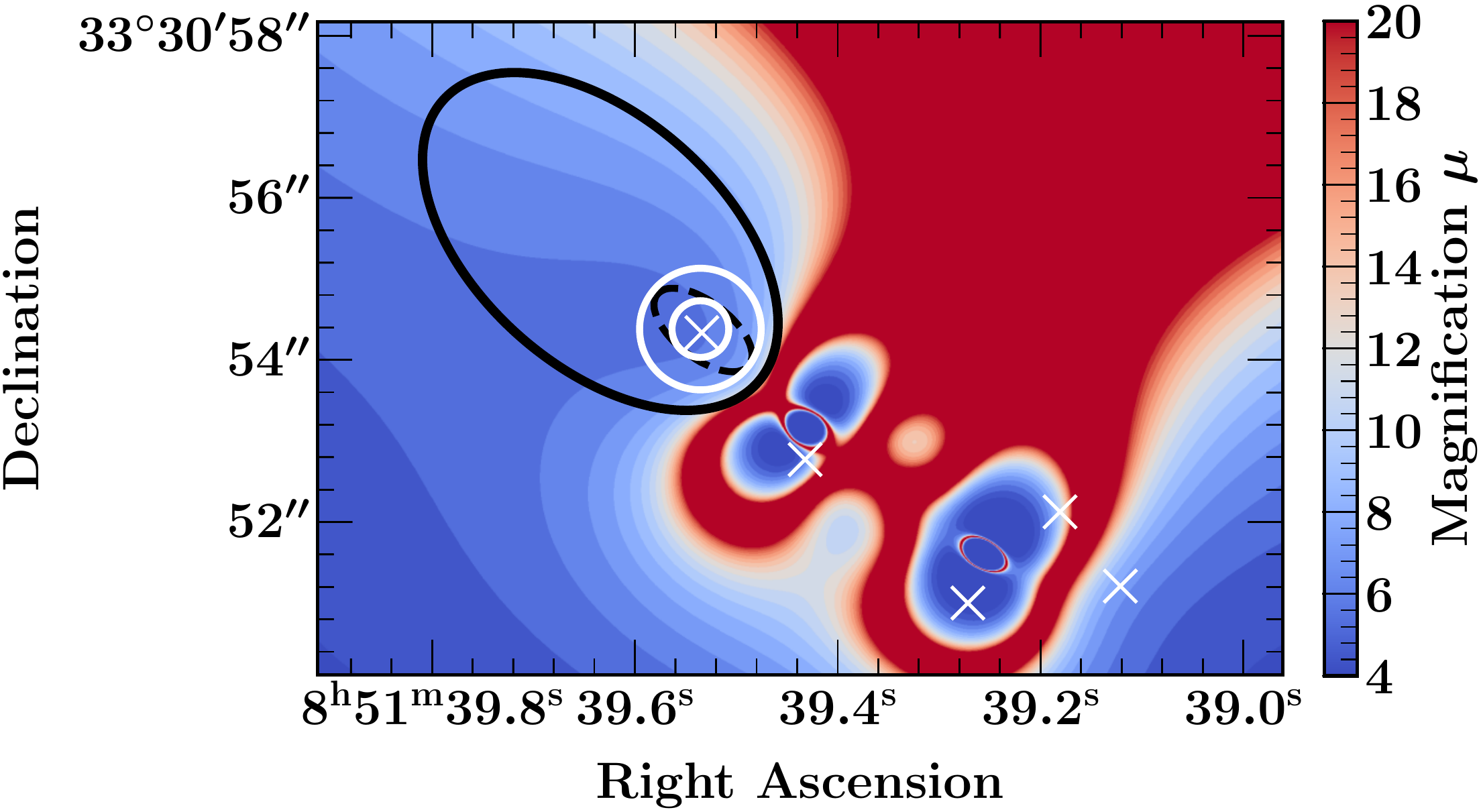}{0.57\textwidth}{}}
\vspace{-0.25in}
\caption{\textbf{Top panel:} Composite color of SDSS~J0851+3331, from \emph{HST}/WFC3 $\mathrm{H_{F160W}}$ (red), $\mathrm{I_{F814W}}$ (green) and $\mathrm{U_{F390W}}$ (blue). The red lines show the critical curves of the gravitational potential at the redshifts of the lensed object $z=1.88$. The cyan circles mark the constraints used to model the system as described in Section \ref{sect:lensmethod}. \textbf{Bottom left panel:} Zoom-in on a region where images of MRG-S0851 appear. The five images are labeled with E1-5.  The green box around image E3 shows the region where the magnification is computed to be $\mu=5.7^{+0.4}_{-0.2}.$ \textbf{Bottom right panel:} The magnification map in the region containing the lensed images of MRG-S0851. The solid black ellipse outlines the segmentation map, and the dashed black ellipse shows the half-light radius. Magnification is almost constant across image 3; $\mu = 6.2^{+1.2}_{-0.7}$ in the solid black ellipse (segmentation map) and $\mu = 5.7^{+0.7}_{-0.2}$ in the dashed black ellipse (half-light radius). The white circles show the $0\farcs7$ and $1\farcs5$ apertures that we adopted to measure global photometry. We indicate the location of the different MRG-S0851 images with crosses.
\label{fig:lens-model}}
\end{figure*}

While we provide a brief summary of the gravitational lensing analysis used in this work here, we refer the reader to \citet{kneib1996}, \citet{richard2010}, \citet{verdugo2011}, and \citet{smith2015} for a more in depth discussion of the lensing algorithm used here. We adopt a parametric approach using \texttt{Lenstool} \citep{jullo2007} to model the cluster mass distribution surrounding our target as a combination of dual pseudo-isothermal ellipsoids \citep[dPIEs,][]{eliasdottir2007}, using a Monte Carlo Markov Chain method to estimate the parameters and their uncertainties. dPIE clumps are combined to map the dark matter (DM) at the cluster scale and to model the cluster mass distribution, and galaxy scale DM potentials are used to describe galaxy scale substructure. Given the large number of galaxies in the cluster, it is not feasible to optimize the parameters of every potential, as the large parameter space will lead to an unconstrained minimization. Moreover, individual galaxies do not contribute significantly to the total mass budget of the cluster, and their effects on lensing are minimal unless they are projected close to the lensed galaxies. To reduce the overall parameter space we therefore scale the parameters of each galaxy using a reference value with a constant mass-luminosity scaling relation \citep[see][]{limousin2007}.

We construct a galaxy cluster catalog using the red sequence technique \citep{gladders2005}, where we select galaxies that have similar colors in the $\mathrm{I_{F814W}}$-$\mathrm{J_{F125W}}$ color versus $\mathrm{J_{F125W}}$-band magnitude diagram. Our original catalog includes 136 cluster members. As the bright cluster galaxies (BCGs) of galaxy clusters do not follow the red sequence, we remove the two BCGs \citep{newman2013,newman2013a} from the galaxy catalog and model them separately. To allow for estimation of lensing magnification, we also decide to remove from the galaxy catalog the two cluster members responsible for the main perturbation of the lensed object and model them separately, too. Additionally, an intrinsic scatter which is expected in the mass luminosity-relation offers further physical motivations to individually model the galaxies by including enough constraints.

Our final lens model of MRG-S0851 includes two cluster-scale DM halos parameterized as dPIE profiles. For minimization procedures, we let all the parameters of halo vary with the exception of the truncation radius $r_{\mathrm{cut}}$ that extends beyond the strong lensing regime, and it therefore cannot be constrained. The final positions of the DM clumps remain close to their BCG ($\leq$5\farcs4). We also tested a model with only one cluster scale DM halo, but the result was not as good as two cluster scale DM halos (see Section \ref{subsec:magnification})

We constrain the cluster using the 3 lensed systems that have confirmed spectroscopic redshifts from \citet{sharon2020} (Figure \ref{fig:lens-model}, left panel). In the two of lensed systems, we identify resolved emission knots withing their image and use them as additional constraints. Figure \ref{fig:lens-model} shows the position of the constraints (see Table \ref{tab:mul-img} for the exact coordinates).
\begin{table}
\begin{center}
\begin{tabular}{|cccccc|}
\hline
ID & Right Ascension & Declination & Notation in \citet{sharon2020} & z & z reference \\
\hline\hline
1.1 & 8:51:38.0254 & +33:31:03.132 & B1.1 & 1.3454 & \citet{bayliss2011}\\
1.2 & 8:51:37.9667 & +33:31:07.032 & B1.2 & 1.3454 & \citet{bayliss2011}\\
1.3 & 8:51:39.4200 & +33:31:26.034 & B1.3 & 1.3454 & \citet{bayliss2011}\\
1.4 & 8:51:39.0580 & +33:31:03.910 & B1.4 & 1.3454 & \citet{bayliss2011} \\
1.5 &  8:51:38.9859 & +33:31:04.751 & B1.5 & 1.3454 & \citet{bayliss2011} \\
2.1 & 8:51:38.0315  & +33:31:02.708 & B2.1 & 1.3454 & \citet{bayliss2011}\\
2.2 & 8:51:37.9613 &  +33:31:07.675 & B2.2 & 1.3454 & \citet{bayliss2011}\\ \hline 
3.1 & 8:51:38.1625 & +33:31:18.666 & D3.1 & 1.79 & this work\\
3.2 & 8:51:38.0261 & +33:30:54.162 & D3.2 & 1.79 & this work\\
3.3 & 8:51:39.5185 & +33:31:03.890 & D3.3 & 1.79 & this work\\
3.4 & 8:51:39.4513 & +33:31:23.948 & D3.4 & 1.79 & this work\\
3.5 & 8:51:38.9480 & +33:31:08.315 & D3.5 & 1.79 & this work\\ \hline 
4.1 & 8:51:39.6640 & +33:30:47.657 & E2.1 & 1.88 & this work\\
4.2 & 8:51:39.6045 & +33:30:46.741 & E2.2 & 1.88 & this work \\
4.3 & 8:51:40.0173 & +33:30:49.868  & E2.3 & 1.88 & this work \\
5.1 & 8:51:39.6474 & +33:30:47.418 & E1.2 & 1.88 & this work  \\
5.2 & 8:51:39.6168 & +33:30:46.898 & E1.1 & 1.88 & this work \\
5.3 & 8:51:39.7549 & +33:30:46.532 & E1.4 & 1.88 & this work \\
5.4 & 8:51:40.0323 & +33:30:49.980 & E1.3 & 1.88 & this work\\
5.5 & 8:51:39.9154 & +33:30:48.302 & E1.5 & 1.88 & this work\\
\hline

\end{tabular}
\caption{Position of the constraints used to construct the lensing models. The coordinates are reported in Sexagesimal coordinates (adopting J2000 epoch). We note that constraints 4.1-4.3 and 5.1-5.5 in this table are labeled as Figure~\ref{fig:lens-model}. We measure the redshifts of each constraint in systems D and E individually, using the default redshift fitting methods of \texttt{Grizli}. We use 12 orbits of grism WFC3/G141 data, obtained as a part of the program, HST-GO-14622, PI: K. Whitaker, for fitting the redshifts of systems D and E. Two constraints 3.2 and 3.4 have clean grism spectra from the program, HST-GO-14622, and the redshift fitting with the grism data was inconclusive for the rest of the constraints 3.1, 3.3 and 3.5 due to contamination of their spectra by nearby objects. The redshift fit of constraints 3.2 and 3.4 is driven both by [OIII] emission-line doublet at a rest-frame wavelength of 5008$\mathrm{\AA}$ yielding a reduced $\chi^2$ of 0.97 and 0.94, respectively.  \label{tab:mul-img}}
\end{center}
\end{table}

\subsection{Modeling Results, Choice of the Best Model\label{subsec:magnification}} 

To estimate a reliable magnification for MRG-S0851, we try several different models: we change the number of cluster scale DM halos, and we also add a fifth image, denoted as system 5.5 (see Table \ref{tab:lens-model-results}). We quantitatively compare the quality of different models with two criteria. The first criterion is the root mean square (rms), which describes how well the model reproduces the positions
of the constraints. The second criterion is the Bayesian Information Criterion (BIC), which is a statistical measurement based on the model likelihood, penalized by the number of free parameters and the number of constraints \citep[see e.g.,][]{Limousin2010,Mahler2018}. 
We list the results in Table \ref{tab:lens-model-results}. Models with two DM halos perform significantly better than the models with one DM halo based on the rms criterion. This effect can be explained by the additional flexibility. However, as the BIC does not change significantly, we conclude that it is balanced by the increase in goodness of the fit for all 2 DM models. The only exception is the 2 DM model with the fifth image as a part of system 4, and we therefore reject it. Finally, we reject the model that shows the best rms (the lowest) and best BIC (the lowest) because the mass of the nearby cluster member galaxy (the one located at the east in Figure \ref{fig:lens-model}, bottom left panel) is unnaturally low. We therefore keep the second best model, with 2 DM and the fifth image as part of system 5 as the best physical model. 

\begin{table}
\begin{center}
\begin{tabular}{|cccccc|}
\hline
Model number & Cluster scale & Fifth image  & rms & BIC & \textbf{$\mu$}\\
{} & DM halo &  assumption &  & & \\
\hline\hline
1 & 2 DM & part of system 5  & 0\farcs21 &  87 & $5.72^{+0.36}_{-0.2}$\\ 
2 & 2 DM & not used & 0\farcs19 & 84 & $5.96^{+0.22}_{-0.19}$\\  
3 & 2 DM & part of system 4 & 0\farcs29 &  96 & $8.45^{+0.9}_{-0.57}$ \\   
4 & 1 DM & part of system 5 & 0\farcs37 &  84 & $8.38^{+0.62}_{-0.46}$ \\   
5 & 1 DM & not used & 0\farcs32 & 77 & $6.45^{+0.22}_{-0.19}$\\    
6 & 1 DM & part of system 4 & 0\farcs36 &  84 & $5.12^{+0.23}_{-0.19}$\\   
\hline

\end{tabular}
\caption{This table summarizes the different modeling assumption and the criteria that we use to compare them. The rms describes how well the model reproduces the positions of the constraints. The Bayesian Information Criterion (BIC) is a statistical measurement based on the model Likelihood, penalized by the number of free parameters and the number of constraints (see e.g., \citealt{Limousin2010}). The last column denotes the estimated magnification, $\mu$, for each model within the green box in the bottom left panel of Figure \ref{fig:lens-model}. \label{tab:lens-model-results}}
\end{center}
\end{table}

Using our final model, we compute the average gravitational magnification in a representative area of the galaxy  to be $\mu=5.7^{+0.4}_{-0.2}$ within the green box in the bottom left panel of Figure \ref{fig:lens-model}. The uncertainty denotes the 1$\sigma$ width of the distribution of $\mu$ for all pixels in the box. We also calculate the magnification in the same box for all models in Table \ref{tab:lens-model-results}.

As can be seen in Table \ref{tab:lens-model-results}, different models lead to different median values of the magnification, and the scatter of the median values for different models is greater than the statistical uncertainty of the best model. We also calculated the light-weighted magnification, the average value of the magnification for all pixels of E3 weighted by its light profile, and we find a mean and a standard deviation of $\mu _l= 7.6 \pm 2.1$ for all models. However, as we discussed earlier, these models can be ruled out based on their physical prediction for the mass of nearby cluster member and/or combination of the rms and BIC. We therefore suggest that the scatter in the median value of all models and the light-weighted magnification is an overestimate, and we use the 1$\sigma$ width of the distribution of $\mu$ for all pixels in the representative area as an estimate of uncertainty in the rest of the paper.

\citet{sharon2020} also present a lensing model for the cluster SDSS~J0851+3331. We note that while \citet{sharon2020} optimize the global properties of the cluster and source positions, we focus on reproducing the geometry of one specific system E, our main science target.  In this work, we construct 6 different models (summarized in Table \ref{tab:lens-model-results}) to study and understand this specific system better. Specifically, the main difference between the models presented herein and \citet{sharon2020} is the added flexibility in the vicinity of the lensed system E. Here, we allow all of the parameters of the galaxies near the systems to vary except for their $x$ and $y$ positions, while \citet{sharon2020} fix $x$, $y$, $e$, and $\theta$ to their observed properties. \citet{sharon2020} also model the core of the cluster with one dominant cluster-scale halo, whereas we find statistical evidence that modeling the core with two halos is favorable.

\section{MRG-S0851 Morphology} \label{sec:whole arc}

In this section, we present our method to constrain the light profile of MRG-S0851 and measure its effective radius. We focus our analyses on E3, the third image (see Figure \ref{fig:0851-E-mosaics}), because it is the brightest image ($m_{\mathrm{H}_{\mathrm{F160W}}}=20.36$) and the grism spectrum is less contaminated by nearby objects. Additionally, E1 and E2 are partially lensed images (Figure \ref{fig:srcrec-GALFIT}), so their light profiles are not a representative of the whole system and henceforth ignored.

\begin{figure*}
\centering
\includegraphics[width=1.0\textwidth]{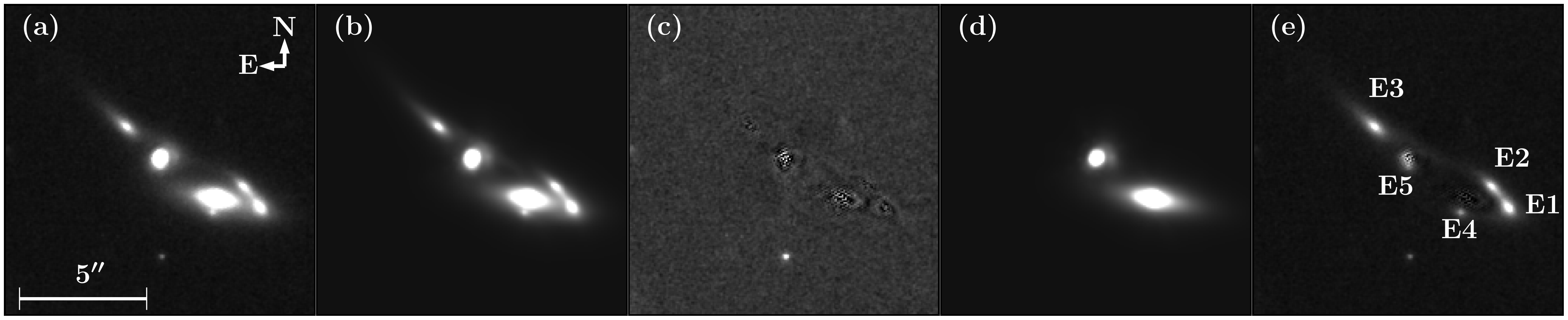}
\caption{(a) The cutout image of MRG-S0851 with the nearby sub-cluster members. (b) The \texttt{GALFIT} model of all light profiles in the field. (c) The residual of the \texttt{GALFIT} model of all light profiles. (d) The \texttt{GALFIT} model of the sub-cluster members (e) The quintuple images of MRG-S0851 after removing sub-cluster lenses using their \texttt{GALFIT} model. The five multiple images of MRG-S0851 are labeled in panel (e). \label{fig:0851-E-galfit}}
\end{figure*}

\subsection{The Image-plane \texttt{GALFIT} Model \label{subsec:size}}

To study the size and the morphology of the different images of MRG-S0851, we use the PSFs and the weight maps that are generated as parts of the data reduction and the photometric analysis together with the $\mathrm{H}_{\mathrm{F160W}}$ drizzled mosaic as basic inputs in the \texttt{GALFIT} software \citep{peng2002}. MRG-S0851 is close to sub-cluster members on the sky, with a significant overlapping of the light profiles. We therefore model out all light profiles in a $12^{\prime\prime}\times 12^{\prime\prime}$ box around the center of the sub-cluster members, using an iterative method to improve our model. The \texttt{Source Extractor} catalog coordinates are used as initial values for the center of the light profiles, the initial S\'ersic index value is $n = 4$, and we use the \texttt{Source Extractor} catalog magnitudes as our initial guess. We run \texttt{GALFIT}, inspect the residuals by eye and add fainter components in the location with the highest residuals. We run \texttt{GALFIT} including the extra components, keeping the components only if there is an improvement in the reduced $\chi ^2$. We fit 14 components in total, fixing the S\'ersic index of two components as well as the axis ratio of a single component in our final iteration to avoid numerical instabilities. None of these components constitute the \texttt{GALFIT} model of E3, and as these components are $\sim$1 ABmag fainter than the \texttt{GALFIT} components of E3, it is unlikely that fixing these parameters affects the \texttt{GALFIT} model of E3 noticeably.  The final result of the image-plane analysis is shown in Figure \ref{fig:0851-E-galfit}. We use the same procedure for the other four \emph{HST} images, obtaining the \texttt{GALFIT} model for all five \emph{HST} filters.

The \texttt{GALFIT} model of E3 includes two components with  S\'ersic indices of $n_1=2.35 \pm 0.05$ and $n_2=1.32 \pm 0.06$, $\mathrm{H_{F160W}}$ magnitudes of $m_1=21.20 \pm 0.03$ and $m_2=21.41 \pm 0.03$, the image plane semi-major axes of $r_1=0\farcs43 \pm 0\farcs01$ and $r_2=2\farcs72 \pm 0\farcs15$, and the axis ratios of $ar_1=0.299\pm 0.003$ and $ar_2=0.150\pm 0.004$ (uncertainties are reported directly from \texttt{GALFIT}). One of these components corresponds to the extended tail of E3. These components are almost aligned with $\delta \theta \sim 9^\circ$. We estimate the spatial offset of 1.3 kpc in the source plane between these two components. The second extended, offset component is not required by the \texttt{GALFIT} model for any of the other \emph{HST} filters, aside from $\mathrm{H_{F160W}}$. We therefore suspect that this offset is a modeling artifact and not physical, caused by non-trivial gravitational lensing effects. 

\begin{figure}
\centering
\includegraphics[width=0.5\textwidth]{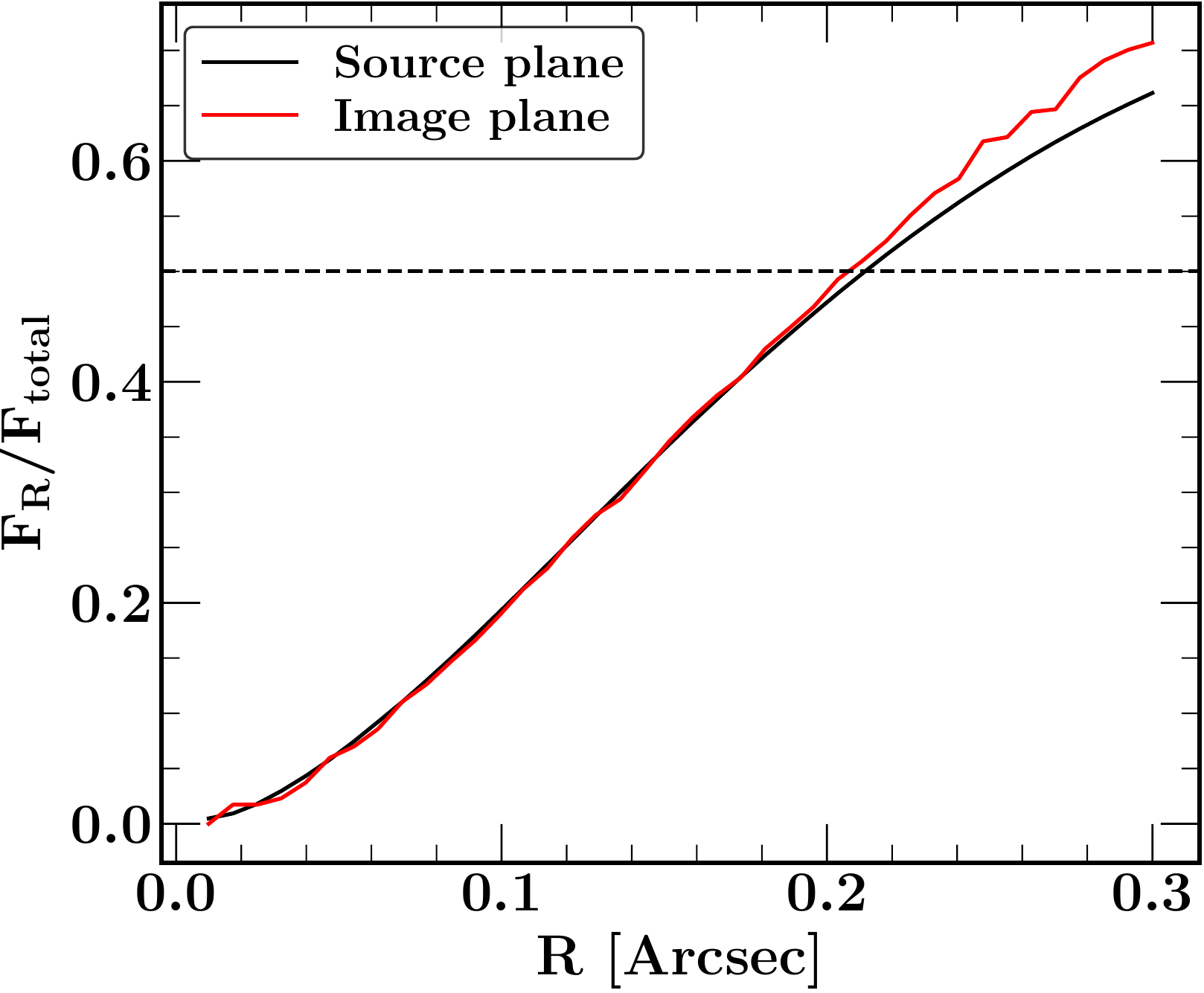}
\caption{The ratio of the enclosed flux to the total flux of PSF-deconvolved images, measured using circular apertures in the source plane and the equivalent elliptical apertures in the image plane. Both ratios are reasonably consistent for different radii of interest, yielding the same half-light radius in both source- and image-plane measurements.}
\label{fig:source-image-comparison}
\end{figure}

\subsection{The Source-plane Effective Radius}

We next measure the effective radius in the source plane using the lensing model (see Appendix \ref{sec:cluster-lens}, for details of the lensing model). To isolate the effect of PSF in our source-plane measurements, we generate a PSF-deconvolved model for E3 and add the residuals, i.e., we calculate $\mathrm{Data}$-$\mathrm{Model_{PSF-convolved}}$+$\mathrm{Model_{PSF-deconvolved}}$   \citep[see][for further discussions of this method]{szomoru2012,newman2018}. After mapping the resulting image back to the source plane, we define a set of circular apertures in the source plane with increasing radii and calculate the corresponding elliptical apertures in the image-plane. We use these two equivalent sets to measure the enclosed flux of the PSF-deconvolved images in both the source and image planes, obtaining a consistent result for the half-light radius, as can be seen in Figure \ref{fig:source-image-comparison}. To measure the uncertainty of the half-light radius, we bootstrap 400 times and remeasure the half-light radius, and in each iteration, we randomize the $\mathrm{H_{F160W}}$-band pixels using the \emph{HST} weight-map uncertainty. The circularized effective radius of E3 is constrained to be $r_c=1.7^{+0.3}_{-0.1}$ kpc ($0\farcs21 \pm 0\farcs02$).

\begin{figure}
\centering
\includegraphics[width=0.5\textwidth]{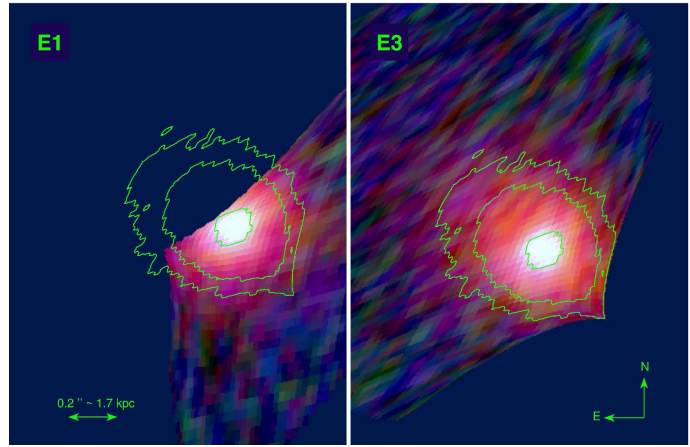}
\caption{Source-plane reconstruction using the \texttt{GALFIT} model presented in Section \ref{subsec:size}. We draw contours from image 3, and over plot them on image 1 after a simple shift. This figure shows that E1 is a partial image.
\label{fig:srcrec-GALFIT}}
\end{figure}

\end{document}